%% file: main.tex
\documentclass[11pt]{article} 
\usepackage[utf8]{inputenc}
\usepackage{geometry}
\usepackage{subcaption}
\usepackage{mathtools} 
\usepackage{fancyvrb}
\usepackage{listings}
\usepackage{enumitem}
\usepackage{textcomp}
\usepackage[toc,page]{appendix}

\usepackage{amssymb}
\usepackage{graphicx}
\usepackage{hyperref}

\title{GenTune: Toward Traceable Prompts to Improve Controllability of Image Refinement in Environment Design}

\author{
  Wen-Fan Wang\textsuperscript{1}\thanks{Both authors contributed equally.} \quad
  Ting-Ying Lee\textsuperscript{1}\footnotemark[1] \quad
  Chien-Ting Lu\textsuperscript{1} \quad
  Che-Wei Hsu\textsuperscript{1} \\
  Nil Ponsa Campanyà\textsuperscript{1} \quad
  Yu Chen\textsuperscript{1} \quad
  Mike Y. Chen\textsuperscript{1} \quad
  Bing-Yu Chen\textsuperscript{1} \\
  \\
  \textsuperscript{1}National Taiwan University, Taipei, Taiwan \\
}

\date{} 

\begin{document}
\maketitle
\renewcommand{\thefootnote}{\fnsymbol{footnote}}
\footnotetext[0]{\hspace{-1.8mm}This is a preprint of the paper accepted at UIST 2025. 
The final version will be available in the ACM Digital Library.}
\renewcommand{\thefootnote}{\arabic{footnote}}
\input{sections/00_Abstract}


\begin{figure}[h]
  \includegraphics[width=\textwidth]{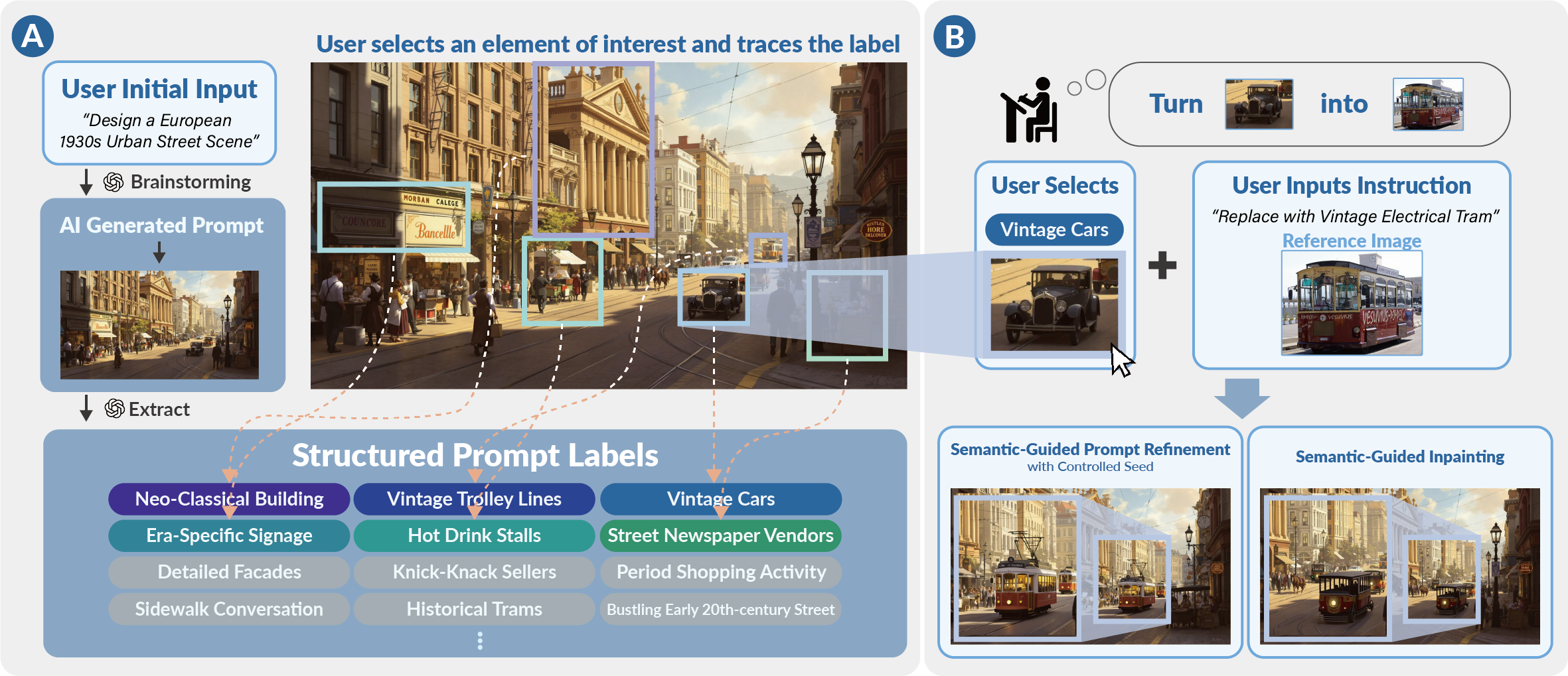}
  \caption{GenTune, a human-centered generative AI system with traceable prompts for controllable image refinement in environment design. (A) The process begins with a user’s initial input-“Design a European 1930s Urban Street Scene”-which is expanded by a Brainstorming LLM into a structured prompt for image generation. A label extraction LLM then extracts key elements to establish prompt–image element correspondences, supporting user understanding. (B) During refinement, the user selects a region of interest to reveal its associated label-“Vintage Cars”. Upon entering a refinement instruction-“Replace with Vintage Electrical Tram”—along with a reference image, GenTune applies both semantic-guided prompt refinement with controlled seed and semantic-guided inpainting to generate updated results.}
  \label{fig:hero image}
\end{figure}


\maketitle

\input{sections/01_introduction}

\input{sections/02_relatedworks}

\input{sections/03_formative_study}

\input{sections/04_systemdesign}

\input{sections/05_summative}
\input{sections/06_results}
\input{sections/07_fieldstudy}
\input{sections/08_discussion}
\input{sections/09_conclusion}

\section*{Acknowledgments}
This work was supported by the National Science and Technology Council, Taiwan (NTSC 112-2221-E-002-185-MY3) and the Center of Data Intelligence: Technologies, Applications, and Systems at National Taiwan University (113L900901, 113L900902, 113L900903), funded through the Featured Areas Research Center Program under the Higher Education Sprout Project by the Ministry of Education (MOE) of Taiwan. We also acknowledge support from National Taiwan University, Moonshine Studio, Winking Studios, and Rayark Games. Finally, we extend our gratitude to all participants and reviewers for their valuable feedback.

\bibliographystyle{plain} 
\bibliography{bibliography}

\appendix

\input{sections/Appendix}

\end{document}

%% file: sections/00_Abstract.tex
\begin{abstract}
Environment designers in the entertainment industry create imaginative 2D and 3D scenes for games, films, and television, requiring both fine-grained control of specific details and consistent global coherence. Designers have increasingly integrated generative AI into their workflows, often relying on large language models (LLMs) to expand user prompts for text-to-image generation, then iteratively refining those prompts and applying inpainting. However, our formative study with 10 designers surfaced two key challenges: (1) the lengthy LLM-generated prompts make it difficult to understand and isolate the keywords that must be revised for specific visual elements; and (2) while inpainting supports localized edits, it can struggle with global consistency and correctness. 
Based on these insights, we present GenTune, an approach that enhances human–AI collaboration by clarifying how AI-generated prompts map to image content. Our GenTune system lets designers select any element in a generated image, trace it back to the corresponding prompt labels, and revise those labels to guide precise yet globally consistent image refinement. In a summative study with 20 designers, GenTune significantly improved prompt-image comprehension, refinement quality and efficiency, and overall satisfaction (all p < .01) compared to current practice. A follow-up field study with two studios further demonstrated its effectiveness in real-world settings.

\end{abstract}

%% file: sections/01_introduction.tex
\section{INTRODUCTION}
\label{sec:introduction}

Environment designers in the entertainment industry craft the visual and spatial worlds that audiences experience in games, animations, films, and TV shows~\cite{bigbadWorld2015, randomguidebook2023, artstation2024, artofgame2008}. Their work typically occurs during pre-production, where they collaborate with art directors to develop 2D and 3D concepts that define a project’s visual direction~\cite{artofgame2008}. These designs often serve as blueprints for modeling and VFX teams, or in smaller or stylized productions, are used directly in final scenes~\cite{domestika2022, cgspectrum2024}. The traditional workflow involves two phases: early ideation—researching, brainstorming, and drafting variations—then final refinement, where approved concepts are polished into production-ready assets~\cite{bigbadWorld2015, conceptart2018, randomguidebook2023}.

With the rise of generative AI (GenAI) tools, environment designers increasingly integrate them into their workflows~\cite{filmhandbook2024, longo2024elaborating, long2024sketchar, wang2025aideation}. They collaborate with multimodal large language models (MLLMs) to craft and refine prompts, which serve as input for text-to-image (T2I) models. 
GenAI is used across ideation, inspiration, client communication, and even production-level outputs~\cite{wang2025aideation, pise2024ai, eisenmann2025expertise}. This shift has raised industry expectations—designers are expected to iterate faster and deliver higher-quality visuals. In response, recent research has explored deeper GenAI integration, such as prompt-tuning~\cite{PromptCharm2024, wang2023reprompt, brade2023promptify} and multimodal models for rapid ideation~\cite{DesignPrompt2024, wang2025aideation, cai2023designaid}.

However, as designers move to refinement, they often need to modify or build upon initial AI-generated outputs. Current GenAI tools treat prompts as opaque, one-shot inputs, making it hard to trace which parts correspond to specific visual elements. While MLLMs like ChatGPT show promise in image editing and structural coherence, 
and their automated nature limits designer control and restricts expressive intent. As a result, designers struggle with precision in the refinement process, often resorting to time-consuming trial-and-error cycles ~\cite{tang2024s, brisco2023exploring, beyan2023review, wang2025aideation}.
In this work, we take a human-centered approach to support understanding and refinement of both AI-generated intermediate prompts and final outputs, tailored to the specific needs of environment designers in the pre-production process.

To understand the challenges environment designers face, we conducted a formative study with 10 participants. Through workflow analysis and in-depth interviews, we identified two core challenges: a lack of understanding of how intermediate text prompts relate to generated image elements, and limited control during the refinement process.
Environment design involves complex spatial and visual composition across both macro (layout) and micro (detail) scales. 
Designers often rely on large language models (LLMs) to generate lengthy, detailed prompts that guide the initial image generation process. They begin with global refinement, adjusting the entire image, including layout or composition. However, for local refinement, they frequently struggle to trace how specific parts of the prompt influence the visual elements, making refinement a trial-and-error process. While inpainting~\cite{yu2018generative} enables targeted edits, it often introduces inconsistencies in lighting, style, or context. More technical solutions, such as ComfyUI~\cite{comfyui} with ControlNet~\cite{zhang2023adding}, offer finer control but are too complex and misaligned with designers' workflows.



To address this, we present GenTune, a human-centered GenAI system designed to enhance prompt-to-image interpretability and the refinement process. For the initial image generation, GenTune builds on prior work~\cite{wang2025aideation, cai2023designaid} by incorporating an MLLM module that expands simple user inputs into detailed prompts. GenTune introduces two key modules following the initial generation: 
\begin{enumerate}
    \item \textbf{Traceable Prompt:} Designers can select an area of interest in the image to reveal a corresponding label traced from the expanded prompt used to generate the image (Fig. ~\ref{fig:hero image}-A). The full prompt segment associated with the label is also displayed, helping designers understand the prompt-image relationship.
    \item \textbf{Semantic-Guided Refinement:} GenTune allows designers to precisely refine the visual elements based on a selected area of interest (Fig. ~\ref{fig:hero image}-B). The system supports three refinement modes: refining the visual element associated with the selected label, modifying only the selected region, or comparing both and choosing the preferred result. Designers can input refinements via natural language and reference images. Additionally, the system suggests refinement options based on both the selected element and the overall image context. 
\end{enumerate}


We conducted a summative study with 20 participants (15 professionals, 5 design students), all experienced with GenAI tools. Across both a controlled experiment and an open-ended task using their own projects, participants rated GenTune significantly higher than a baseline that lacks GenTune's two key modules in prompt–image understanding (\textit{p} = 0.003), refinement effectiveness (\textit{p} = 0.002), output quality (\textit{p} = 0.003), and overall satisfaction (\textit{p} < 0.001). Compared to their typical workflows, they also preferred GenTune for satisfaction, efficiency, quality, and creativity support (\textit{p} < 0.001 for all).
Following the summative study, we conducted a field study with two design studios to evaluate GenTune in real-world professional settings. Designers integrated GenTune into their ongoing commercial projects for three days. All participants reported improved efficiency, higher output quality, and a significant reduction in time spent communicating design intent to clients and directors.

In summary, the major contributions of this work are:
\begin{itemize}
    \item A formative study investigating the end-to-end GenAI workflow of professional environment designers, identifying key challenges and specific needs in refining generated images.
    \item 
    The design and implementation of GenTune, a human-centered GenAI system that supports targeted, semantic image refinements via natural language or image references.
   \item A HCI paradigm for traceable, element-level control, enhancing understanding and controllability in human–AI collaboration.
    \item A comprehensive multi-stage evaluation showing that GenTune significantly improves designers’ ability to interpret, control, and refine generative outputs. This includes: (1) a within-subjects experiment, (2) an open-ended task using production projects, and (3) a field deployment in two design studios.
\end{itemize}

%% file: sections/02_relatedworks.tex
\section{RELATED WORK}
\subsection{Generative AI in Creative Workflow} 
The creative process involves both divergent and convergent thinking. Designers begin by exploring a wide range of possibilities to generate diverse ideas~\cite{wu2016imagination, kim2013convergent, zhu2019convergent, xie2023cognitive}, and iteratively refine selected concepts through evaluation cycles~\cite{johnson1997analysis, stigliani2018shaping, javaid2021association}. This process often requires a non-linear, iterative process between exploration and refinement ~\cite{adams1999cognitive, chong2025prompting, frich2021digital}.

Previous work has leveraged GenAI to support divergent stages of creativity, aiding visual exploration and early ideation. These systems help users generate various outputs~\cite{liu2022opal}, explore unexpected directions~\cite{ko2023large}, and have been applied in domains such as fashion~\cite{jeon2021fashionq}, architecture~\cite{tan2024using}, and 3D CAD modeling~\cite{liu20233dall}. For visual exploration, some tools streamline prompt crafting for faster feedback~\cite{liu2022design, chung2021intersection}, while others support conceptual expansion and recombination. For example, CreativeConnect~\cite{choi2024creativeconnect}, PopBlends~\cite{wang2023popblends}, and GenQuery~\cite{son2024genquery} help designers decompose and blend concepts to generate novel ideas. DesignAID~\cite{cai2023designaid} uses LLMs to expand design space, and AIdeation~\cite{wang2025aideation} further enables flexible recombination of retrieved visual references and generated content.

Recent work has integrated generative AI into the creative field to enhance domain-specific workflows and communication throughout the design process. For example, MemoVis~\cite{chen2024memovis} translates textual feedback into visual references to reduce misinterpretation in 3D modeling, while RoomDreaming~\cite{wang2024roomdreaming} enables collaborative spatial exploration through photorealistic interiors. PlantoGraphy~\cite{huang2024plantography} visualizes abstract prompts for iterative landscape design, and Keyframer~\cite{tseng2024keyframer} offers a natural language interface for motion design. Other applications span video co-creation~\cite{huh2025videodiff}, creative writing~\cite{fu2025like}, and spatial design~\cite{wan2024breaking}. These tools demonstrate the potential of GenAI to accelerate exploration and improve communication. 


Unlike most prior systems that focus on divergent idea exploration, GenTune extends these capabilities into the refinement stage—supporting convergent processes and addressing challenges such as maintaining visual consistency, a key challenge in current environment designers’ GenAI workflows.

\subsection{Addressing Understanding in Human-AI Design Collaboration} 

Understanding and effectively interacting with GenAI remains a challenge, especially for non-experts~\cite{kim2023help}. Despite advances in T2I models, users often struggle to clearly express their intention and achieve the desired results~\cite{bousetouane2025generative, zhang2023generative}. A key challenge lies in the unintuitive nature of prompt crafting, which requires careful planning and remains difficult for non-expert users~\cite{zamfirescu2023herding, zamfirescu2023johnny}. While many systems aim to optimize prompts~\cite{wang2024learning, ahmad2024crafting}, manual prompt creation is often still necessary, especially for environment designers, who heavily rely on commercial tools like MidJourney\footnote{MidJourney, https://www.midjourney.com/}. Although some artists and designers use LLMs to assist in prompt generation~\cite{lindley2025prompt, li2024flowgpt}, a gap remains in users’ understanding of T2I models. They often struggle to connect prompts with generated outputs~\cite{oppenlaender2024prompting}, resulting in limited control and reliance on trial and error~\cite{guo2024prompthis}.


To address these challenges, Explainable AI (XAI) approaches aim to make the generation process more transparent and interpretable~\cite{dwivedi2023explainable, bansal2021does, wang2021explanations}. However, much of this work remains system-centered, focusing on algorithmic transparency over user understanding~\cite{kim2023help}. Recent HCI research has begun to bridge this gap with human-centered AI approaches~\cite{Usmani2023Human-Centered, liao2021human, ehsan2022human}, emphasizing user engagement and dynamic interaction~\cite{raees2024explainable, thieme2020interpretability}. This shift prioritizes interpretability as a means to improve user experience and support human-AI collaboration.

For example, From Text to Pixels~\cite{evirgen2024text} uses visual and sensitivity-based explanations to show how text prompts influence outputs in T2I models. Other systems support fine-grained understanding through different modalities: AutoSpark~\cite{chen2024autospark} allows detailed comparisons to improve text-image relevance; XCreation~\cite{yan2023xcreation} offers a graph-based interface for intuitive control in cross-modal story creation; and ProtoDreamer~\cite{zhang2024protodreamer} supports physical prototyping with AI feedback on design evolution. Interactive systems in music~\cite{louie2020novice}, translation~\cite{coppers2018intellingo}, and crowd ideation~\cite{wang2022interpretable} likewise enhance user understanding by exposing AI reasoning and supporting more directed interaction.


Whereas prior works adopt diverse strategies to enhance user understanding of the generative process, GenTune's focus on prompt-to-image element mapping is action-oriented: traceability is not the end goal, but a mechanism to enable direct, element-centric manipulation and refinement.

\subsection{Refinement in Generative Image Workflows} 
While generative AI has been widely studied for its role in the divergent stages of creativity, such as ideation and iterative design~\cite{sun2024generative, palani2024evolving}, fewer studies have examined its potential to support convergent processes, as highlighted in recent research~\cite{rodrigues2023creativity, chiu2024elevating}.

Recent work has supported image refinement through interactive and automated tools. RePrompt~\cite{wang2023reprompt} and Promptify~\cite{brade2023promptify} enable iterative prompt editing, with RePrompt using CLIP-based scoring for alignment.  PromptCharm~\cite{PromptCharm2024} and PromptMagician~\cite{feng2023promptmagician} offer interfaces for exploring and editing prompt variations, with PromptCharm additionally suggesting terms based on user intent. PromptMap~\cite{adamkiewicz2025promptmap} enables semantic navigation across prompt spaces, while DreamSheets~\cite{almeda2024prompting} visualizes prompt exploration to aid discovery and refinement. Fermat~\cite{serra2022programmable} provides multimodal control, supporting both exploratory and refinement phases.
These systems alleviate the trial-and-error burden of prompt crafting by directly manipulating prompts. In contrast, GenTune utilizes LLMs to generate structured prompts from simple input and enables refinement post-generation, thereby bypassing the need for manual prompt crafting.

Beyond prompt-based refinement, recent work has explored image-based techniques—mainly built on Stable Diffusion~\cite{rombach2022high, ruiz2023dreambooth}—to enhance control over generative outputs. LoRA~\cite{hu2021lora} enables lightweight style tuning, while InstructPix2Pix~\cite{brooks2023instructpix2pix} allows edits via natural language. Inpainting methods~\cite{lugmayr2022repaint, yu2023inpaint} support selective region regeneration. ControlNet-based tools~\cite{zhang2023adding}, using inputs like edges, depth, or segmentation maps, provide structural conditioning and are widely adopted in platforms like ComfyUI~\cite{xue2024genagent}. However, these methods often come with steep learning curves~\cite{zhang2024research} and may misalign with creative workflows. Seed-based generation offers an alternative, where consistent seeds and detailed prompts yield stable results~\cite{pronin2024evaluating, xu2024good}, but this typically demands precise prompt engineering~\cite{ge2024seed, li2024enhancing}.


Beyond these methods, multimodal interfaces have been explored to provide more expressive and localized control in generative image workflows. A common approach is sketch-based interaction: SketchFlex~\cite{lin2025sketchflex} refines rough sketches into structured anchors for spatial-semantic coherence, while Inkspire~\cite{lin2025inkspire} supports analogical sketching for early-stage ideation. PromptPaint~\cite{chung2023promptpaint} and Exploring Visual Prompts~\cite{park2025exploring} use scribbles, annotations, and brushes to guide generation. DesignPrompt~\cite{DesignPrompt2024} combines images, colors, and text in a moodboard-style prompt interface. Other systems support human–AI co-creation via multimodal control: GANzilla~\cite{evirgen2022ganzilla} spans both exploratory and refinement phases, while GANravel~\cite{evirgen2023ganravel} enables user-driven direction disentanglement for iterative GAN editing.

More recently, MLLMs like ChatGPT-4o and Gemini 2.0 Flash\footnote{\label{Gemini}Gemini, https://developers.googleblog.com/en/experiment-with-gemini-20-flash-native-image-generation/} enable image editing through dialogue-based interactions, offering prompt enhancement and structural control. While promising, they still fall short in meeting the specific refinement needs of environment design, as discussed later.

While prior work focuses on diverse refinement approaches, few address multi-step, LLM-assisted input generation, where controlling intermediate outputs is critical. GenTune introduces traceable, element-level control to improve understanding and controllability in future MLLM-assisted workflows.


%% file: sections/03_formative_study.tex
\section{FORMATIVE STUDY}
We conducted a formative study to investigate how environment designers currently use GenAI design tools, explore their workflows, and identify the challenges they face.

\subsection{Participants}

We recruited 10 participants, including 5 professional environment designers from the game (P1, P4, P6) and animation (P3, P5) industries (3–8 YoE, mean = 4.8), and 5 design students (P2, P7–P10) with at least six months of intensive GenAI experience. 
Participants were recruited through personal referrals within the local art community and by directly contacting studios by email to request collaboration.
Participants were compensated ~15 \$USD. Demographics, including study participation, years of experience, and commonly used AI tools, are shown in Table~\ref{tab:demographics}.

\subsection{Study Procedure}
The study consisted of three parts. We began with a 20-minute in-depth interview covering: (1) participants’ goals when using GenAI tools, (2) their strategies and workflows, and (3) past experiences using GenAI tools in projects, including challenges they encountered. A 30-minute workflow observation session followed to gain deeper insight into their design strategies. Participants used their preferred GenAI tools to design a Medieval Chinese Science Center. Finally, a 20-minute post-task interview focusing on their refinement strategies and difficulties. Each session lasted 1–1.5 hours. 

\subsection{Findings}
\subsubsection{Data analysis}
We conducted a thematic analysis of summarized and transcribed interviews. An author with professional environment design experience developed the initial coding framework, which was refined with feedback from two art directors overseeing 21 and 34 concept designers. Thematic analysis was discussed collaboratively
among three co-author to ensure consensus.

\subsubsection{Roles of GenAI in Environment Design}

Designers commonly use GenAI tools for early-stage ideation. They begin by using ML models to expand simple prompts and generate images as visual references, especially when working with unfamiliar design specifications (P2-4, P6-9). As one participant noted, “\textit{It's common that we cannot find specific design references to the topic, but GenAI tools can provide that}” (P4). 
As GenAI tools become more widespread, art directors and clients increasingly expect environment designers to use them for faster visual communication and decision making. 
This shift has raised expectations for speed and quality. As one participant described, 
“\textit{Clients think GenAI is powerful and expect high-quality revisions daily, not every few days like before}” (P3). 
Despite this pressure, GenAI output still requires significant manual refinement (P1, P3-6). As one designer explained, “\textit{AI rarely captures exactly what the client wants—we still do a lot of post-processing to meet their expectations}” (P4).

\subsubsection{Current GenAI Refinement Workflow and Key Challenges}
As shown in Table \ref{tab:demographics}, participants used various image generation tools. 
Despite those differences, their workflows were similar: after initial generation, they refined the images iteratively from global structures to local details.
During the refinement, we found designers often emphasize several elements such as theme, art style, composition, lighting, color, and shot angle.

They usually begin with prompt editing. However, a key challenge was the lack of clear mapping between the LLM-generated prompt and visual elements, making it hard to determine which parts of the prompt control specific image features (P2-4, P6-7, P9-10). “\textit{I often have to search for a long time just to find the section I need to edit}” (P10). Even then, designers were unsure if their edits had the intended effect. “\textit{Many times the new image makes me question whether I actually edited the right part}” (P2).
As a result, designers often rely on trial and error. Even with LLM support, ambiguity persists. “\textit{I feel like I’ve described exactly how to change it, but the AI still doesn’t understand}” (P9). 
Moreover, generative images also include unexpected elements not mentioned in the prompt; for these, neither manual edits nor LLM assistance are effective. These challenges underscore the inefficiencies of prompt-based refinement, especially in complex environment design.

Another key issue is structural consistency, which is critical in environment design, especially when presenting to stakeholders. However, general-purpose GenAI tools often fail to preserve structural coherence (P1, P3-5, P7-9). “\textit{If we need to revise an AI image for clients, we usually just edit it manually in Photoshop—it’s faster}” (P4). Some designers tried technical workflows like using structural conditioning in ComfyUI~\cite{comfyui}, but found them ineffective for complex scenes. “\textit{Even if the structure stays the same, all the textures go off}” (P1). Others found these tools too complex to use. As P3 put it, “\textit{I'm an artist, not an engineer—why should I have to use this (ComfyUI)?}”

To modify local details without affecting other regions, some designers used prompt-based inpainting~\cite{yu2018generative}. This involves selecting a region and entering a replacement prompt. However, results were often visually inconsistent—mismatched lighting, style, proportion, or contextual incoherence, such as historical or spatial inconsistency (P1-3, P6-9). “\textit{Inpainting often gives me strange results. I’ve tried many different prompts, but still didn’t get what I wanted}” (P6). Designers emphasized that visual coherence was more important (P1, P3-5, P7-10): “\textit{Although inpainting preserved other areas, the materials were completely wrong. I would rather start over.}” (P7). Others opted for manual editing, but the lack of editable layers made this difficult. “\textit{It’s faster to repaint the area from scratch}” (P3). Still, when many similar elements needed changes, manual editing “\textit{just takes too much time}” (P4).

In summary, environment designers struggle to understand how LLM-generated prompts map to visual outputs and lack support for targeted refinement. They prioritize visual coherence over pixel-level accuracy—something current GenAI tools often fail to provide.

\subsection{Design Goals}
Based on the findings, we proposed three design goals for our system:
\begin{itemize}
    \item \textbf{DG1: Transparent Prompt-to-Image Mapping.} The system should enable clear, interactive mappings between prompt text and corresponding visual elements, helping environment designers understand how specific prompts influence outputs. 
    \item \textbf{DG2: Maintaining Coherence in Refinement.} The system should align with designers’ priorities by maintaining visual and semantic coherence throughout the iterative refinement process, from global structure to local details.
    \item \textbf{DG3: Element-Centric and Predictable Control Workflow.} The system should support element-centric manipulation that offers precise control and predictability, reducing trial-and-error and enhancing designers’ creative exploration.
\end{itemize}



\begin{table*}[h!]
\centering
\small
\renewcommand{\arraystretch}{1.2} 
\begin{tabular}{|c|c|c|p{2.2cm}|p{5cm}|c|c|c|}
\hline
\textbf{ID} & \textbf{Age} & \textbf{YoE} & \textbf{Industry} & \textbf{GenAI Tools Used} & \textbf{Formative} & \textbf{Summative} & \textbf{Field} \\ \hline
1  & 34 & 8  & Game & Midjourney, ComfyUI & \checkmark & & \\ \hline
2  & 22 & 0  & Student & ChatGPT (DALL·E) & \checkmark & \checkmark & \\ \hline
3  & 27 & 3  & Animation & Midjourney & \checkmark & \checkmark & \checkmark \\ \hline
4  & 27 & 4  & Game & Midjourney, ChatGPT (DALL·E) & \checkmark & \checkmark & \checkmark \\ \hline
5  & 27 & 5  & Animation & Midjourney & \checkmark & \checkmark & \\ \hline
6  & 27 & 4  & Game & Stable Diffusion & \checkmark & \checkmark & \\ \hline
7  & 23 & 0  & Student & Midjourney, Stable Diffusion, ChatGPT (DALL·E), Leonardo & \checkmark & \checkmark & \\ \hline
8  & 22 & 0  & Student & Midjourney, ChatGPT (DALL·E), Leonardo & \checkmark & \checkmark & \\ \hline
9  & 22 & 0  & Freelancing, Animation, Student & ChatGPT (DALL·E), ComfyUI & \checkmark & \checkmark & \\ \hline
10 & 23 & 0  & Animation, Student & Midjourney, Stable Diffusion, ChatGPT (DALL·E) & \checkmark & \checkmark & \\ \hline
11 & 27 & 4  & Game & Midjourney & & \checkmark & \\ \hline
12 & 39 & 10 & Game & Midjourney, Stable Diffusion & & \checkmark & \\ \hline
13 & 39 & 13 & Game & ChatGPT (DALL·E) & & \checkmark & \\ \hline
14 & 24 & 1  & Animation & Midjourney & & \checkmark & \\ \hline
15 & 30 & 8  & Freelancing, Film/TV & Stable Diffusion & & \checkmark & \\ \hline
16 & 27 & 5  & Freelancing, Film/TV, Animation & Midjourney, Stable Diffusion, ChatGPT (DALL·E) & & \checkmark & \\ \hline
17 & 43 & 21 & Game & Stable Diffusion, ChatGPT (DALL·E), ComfyUI & & \checkmark & \\ \hline
18 & 28 & 5  & Freelancing & Midjourney, ChatGPT (DALL·E) & & \checkmark & \\ \hline
19 & 31 & 6  & Freelancing & Stable Diffusion & & \checkmark & \\ \hline
20 & 28 & 3  & Industrial Design & Midjourney, Firefly, Vizcom & & \checkmark & \\ \hline
21 & 30 & 7  & Animation & ChatGPT (DALL·E), Moonshot & & \checkmark & \\ \hline
22 & 27 & 3  & Animation & Midjourney, Stable Diffusion & & & \checkmark \\ \hline
\end{tabular}
\caption{Demographic details of participants including age, industry, generative AI tools used, and study participation.}
\label{tab:demographics}
\end{table*}

%% file: sections/04_systemdesign.tex
\section{SYSTEM \& IMPLEMENTATION}
We present GenTune, a human-centered generative AI system that enhances interpretability and user control in image refinement for environment design. GenTune helps designers quickly identify areas of interest and supports precise, progressive refinement.

\begin{figure*}[htbp]  
    \centering
    \includegraphics[width=1\linewidth]{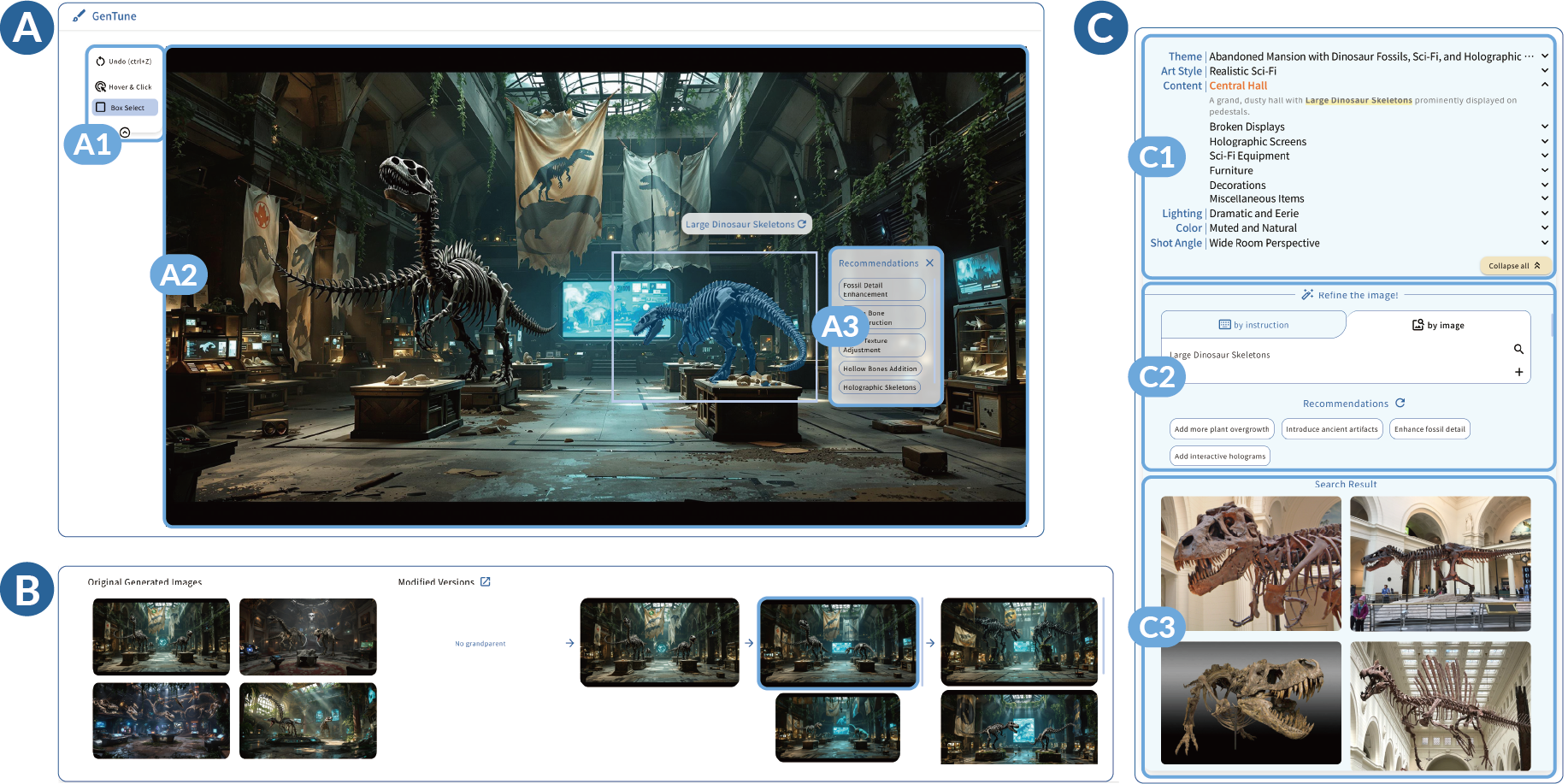}
    \caption{The main interface of GenTune includes: (A) The main image panel. Users can select elements either by hover-and-click or box-selection using the mode selector (A1). Once a region is selected, the image (A2) displays the corresponding prompt label and refinement suggestions (A3). (B) The version history view, showing the initially generated images and the iterative refinement history of the selected image. (C) The refinement panel. (C1) The structured prompt view, the corresponding section expands automatically when a region is selected. (C2) Input dialog with text input suggestions. If “by instruction” is selected, users can enter text commands to refine the image. (C3) If “by image” is selected, users can search for reference images.}
    \label{fig:ui}
\end{figure*}

\subsection{System Overview}
GenTune’s image generation system draws inspiration from conversational generation systems ~\cite{huang2024dialoggen, wei2023dialogpaint, wang2025aideation}, which support iterative, dialogue-driven workflows. It features a brainstorming module that expands simple user input into high-quality prompts and generates four initial images through a conversational interface.

Once an image is selected, it enters GenTune’s main interface (Fig. \ref{fig:ui}), which features two core modules designed to support our key goals.

\subsubsection{Traceable prompt}
GenTune structures prompts into six key categories—theme, art style, content, lighting, color, and shot angle—reflecting the most emphasized aspects of environment design from our formative study.

To help designers identify which parts of the prompt correspond to visual elements, 
GenTune lets designers select an element in the image to reveal a corresponding label traced from the expanded prompt; the related section in the prompt panel then expands automatically (Fig.~\ref{fig:ui}-C1). Labels are organized in a tree structure, with parent categories expanding when traced, and are derived from the "content" category, which maps to identifiable visual elements.

\subsubsection{Semantic-guided refinement}
After tracing a label, designers can refine the image by entering natural language instructions in the dialogue box, or providing a reference image via the embedded search engine or by uploading directly through the right-hand panel (Fig. \ref{fig:ui}-C2).

GenTune supports three refinement options, each operating at a different scope, from global to local:
\begin{itemize}
    \item \textbf{Global refinement (no selection required)}. Designers can make broad, whole-image edits using text or reference images, such as changing style, mood, or lighting, without selecting specific regions.
    \item \textbf{Semantic-guided prompt refinement with controlled seed (requires selection)}. In this novel method, GenTune makes targeted edits to the original prompt based on the designer’s input and the selected label, then regenerates the image using the same seed. This allows changes to apply only to the intended element while preserving overall coherence ~\cite{pronin2024evaluating, ge2024seed}. For example, the designer selects the element labeled “Vintage Cars” and provides a reference image of a vintage tram. GenTune replaces cars with trams and adds overhead wires to ensure contextual consistency, while keeping the rest of the image largely unchanged (Fig. ~\ref{fig:hero image}-B, Fig. \ref{fig:seed_better}-B). This design choice aligns AI-driven refinement with designer intent by prioritizing conceptual consistency and visual coherence over exact pixel-level accuracy.
    \item \textbf{Semantic-guided inpainting (requires selection)} This option enables more localized edits. 
    Unlike traditional inpainting, GenTune accepts simple natural language commands (e.g., “add some merchants”) and generates context-aware prompts using the selected label and original prompt. This allows the inserted content to remain stylistically and semantically consistent with the overall scene.
\end{itemize}

\begin{figure*}
    \centering
    \includegraphics[width=1\linewidth]{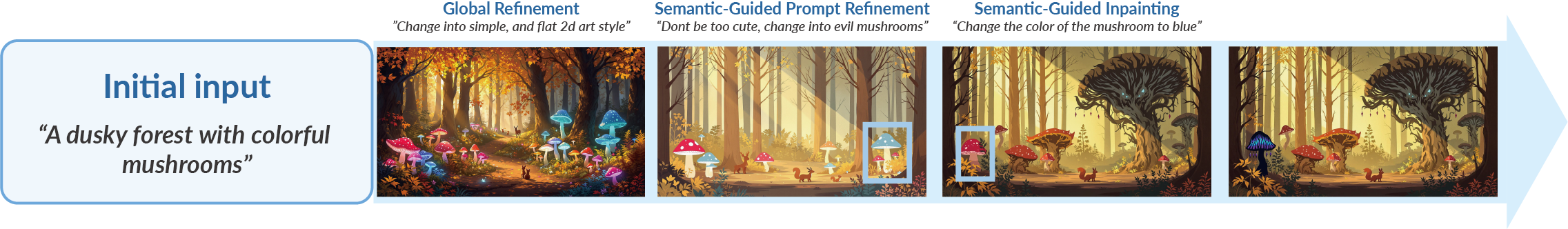}
    \caption{GenTune’s refinement workflow from P14 in the open-ended study began with Global Refinement to adjust the overall style, followed by Semantic-Guided Prompt Refinement to turn all mushrooms into "evil" variants, and concluded with Semantic-Guided Inpainting for localized edits, such as changing the color of a single mushroom. }
    \label{fig:flow}
\end{figure*}

As shown in Figure~\ref{fig:flow}, GenTune supports a progressive refinement workflow that aligns with designers’ natural editing process—starting from global adjustments to fine-grained control.
Users typically begin with Global Refinement for large-scale changes (e.g. art style), followed by Semantic-Guided Prompt Refinement to adjust specific elements while preserving overall structure, and finally Semantic-Guided Inpainting to make precise, localized edits.

For each refinement, GenTune generates four image variations. It offers three modes: seed mode and inpainting mode, each producing four results, and mixed mode, which returns two from each. This allows designers to compare outcomes across strategies, especially useful when they are unsure about the scale of change or want to explore stylistic trade-offs.


GenTune provides three types of refinement suggestions: (1) global suggestions for broad edits like lighting or content (Fig.\ref{fig:ui}-C2); (2) label-based suggestions for element-specific changes (Fig.\ref{fig:ui}-A3); and (3) expanded suggestions that build on user input to offer detailed, context-aware options. These aid designers, especially when unsure of the next step.

\begin{figure}[htbp]
    \centering
    \includegraphics[width=1\linewidth]{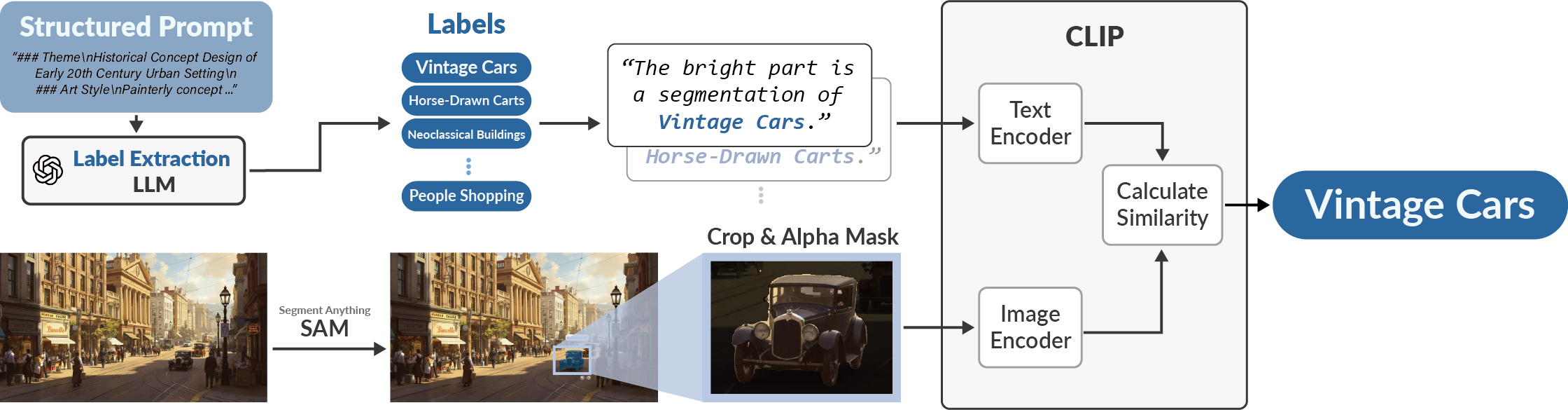}
    \caption{Prompt-element correspondence pipeline.  Label Extraction LLM extracts labels from the structured prompt and used as CLIP's text inputs. When a user selects an element, it is segmented using SAM, cropped with a bounding box, alpha-masked, and passed into CLIP as the image input. CLIP then computes the text-image similarity to determine the label most associated with the selected region.}

    \label{fig:clip}
\end{figure}

\subsection{User Interface}
GenTune features a web-based interface with three main pages: (1) a front page for initial input, (2) an overview page displaying four generated images, and (3) the main interface with GenTune’s two core modules (Fig. \ref{fig:ui}).
In the main interface, users can hover or draw a box to reveal element labels on the image (Fig. \ref{fig:ui}-A2), click the refresh icon to cycle through alternatives, and use the checkmark icon to view label-based suggestions.
The right panel contains the prompt overview (Fig. \ref{fig:ui}-C1) and a dialog box (Fig. \ref{fig:ui}-C2) for entering text or uploading reference images. Users can switch between Mixed, Seed, and Inpainting modes via the mode button. Prompt suggestions appear below and can be refreshed based on the user’s input.
The bottom of the interface displays an image iteration tree (Fig. \ref{fig:ui}-B), showing thumbnails of the hierarchical relationships of each image, to help track and revisit edits.

\subsection{Technical Implementation}
GenTune uses GPT-4o-2024-08-06\footnote{GPT-4o, https://platform.openai.com/docs/models/gpt-4o} as the base language model, and Flux 1.1 Pro Ultra\footnote{Flux 1.1, https://blackforestlabs.ai/ultra-home/} as the image model, for its strong performance with natural language prompts. The average generation time for a single iteration is approximately 30 seconds. The backend and frontend are deployed on a Linux-based personal computer with an Nvidia GeForce RTX 4080 GPU. Upon region selection, the system returned the corresponding label with an average latency of 1 second.

\subsubsection{Traceable prompt}
After the intial input, the system generates a structured prompt with key categories with a Brainstorming LLM (Fig. \ref{fig:hero image}-A). A Label Extraction LLM extracts key elements as labels from the Content category, which then are used to define the class set for subsequent detection (Fig. \ref{fig:clip}).
When a user selects an element—either by hovering and clicking or drawing a box—the system sends a point or box prompt to the SAM model~\cite{kirillov2023segment} to obtain the corresponding segmentation mask and bounding box. These selection methods are designed to support interaction needs, allowing for either quick region selection or more precise control. To enable real-time interaction, we precompute image embeddings using the ONNX run-time version of SAM, deployed on the frontend.

Based on the bounding box, we crop the image and darken 80\% of the area outside the segmentation mask. This cropped image is passed to the CLIP model~\cite{radford2021learning} for semantic similarity matching. For each candidate label in the class set, we use the prompt: “The bright part is a segmentation of {label}” for text encoding. CLIP encodes both the image and these text descriptions into a shared embedding space, and computes similarity scores. The top five labels are returned as the predicted semantic tags for the selected region.

\begin{figure}[htbp]
    \centering
    \includegraphics[width=1\linewidth]{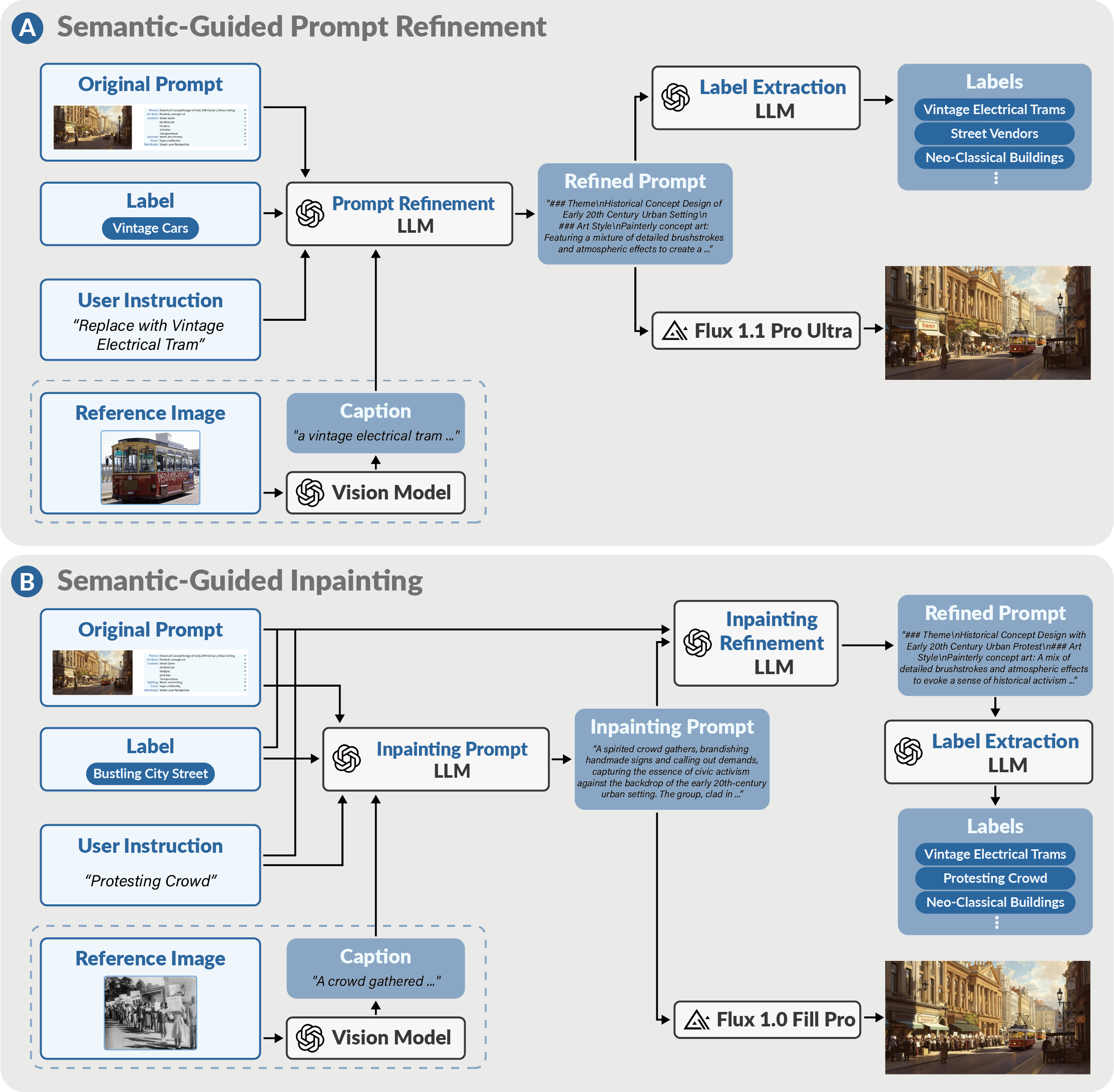}
    \caption{Technical pipeline for semantic-guided prompt refinement and inpainting. If a reference image is provided, it is captioned by a vision model. (A) The Prompt Refinement LLM takes the original prompt, extracted label, user instruction, and optional reference image caption as input to generate a refined prompt for full-scene generation. This prompt is sent to the Flux 1.1 Pro Ultra model to synthesize a new image, and concurrently is passed to the Label Extraction LLM to update semantic labels. (B) The Inpainting Prompt LLM generates a region-specific prompt using the same inputs. This prompt guides the Flux 1.0 Fill Pro model to inpaint the selected region. In parallel, the Inpainting Refinement LLM merges the new original and inpainting prompts to generate a refined prompt for label extraction.}
    \label{fig:refinement}
\end{figure}

\subsubsection{Semantic-guided refinement}

For global refinement, which does not require label selection, the Global Refinement LLM processes user instructions to generate a new structured prompt. 

For semantic-guided prompt refinement (Fig. \ref{fig:refinement}-A), textual instructions are processed by a Refinement LLM, which takes the selected label, the user’s input, and the original prompt as input. It identifies the region of the prompt associated with the label and performs highly precise modifications based on the instruction. For image-based refinement, a vision model generates a caption of the reference image, which is combined with the instruction to update the corresponding prompt segment. The refined prompt is then used to (1) generate a new image with the same seed and (2) extract updated keywords for downstream use.

For semantic-guided inpainting (Fig. \ref{fig:refinement}-B), the Inpainting Prompt LLM takes similar input and generates a detailed prompt. The inpainting prompt is crafted to remain consistent with the original prompt in style and semantics. We use Flux 1.0 Fill Pro, a state-of-the-art inpainting model, to apply the modification directly to the image.
Since inpainting operates directly on the image rather than modifying the prompt, it is not inherently compatible with prompt-based regeneration. To address this, GenTune reconstructs a new structured prompt from the original and inpainting prompts by the Inpainting Refinement LLM. This allows for updated label extraction and optionally enables further prompt-based generation. However, this is not the recommended workflow, as the new generation may overwrite the inpainting edits.

\begin{figure}[htbp]
    \centering
    \includegraphics[width=0.8\linewidth]{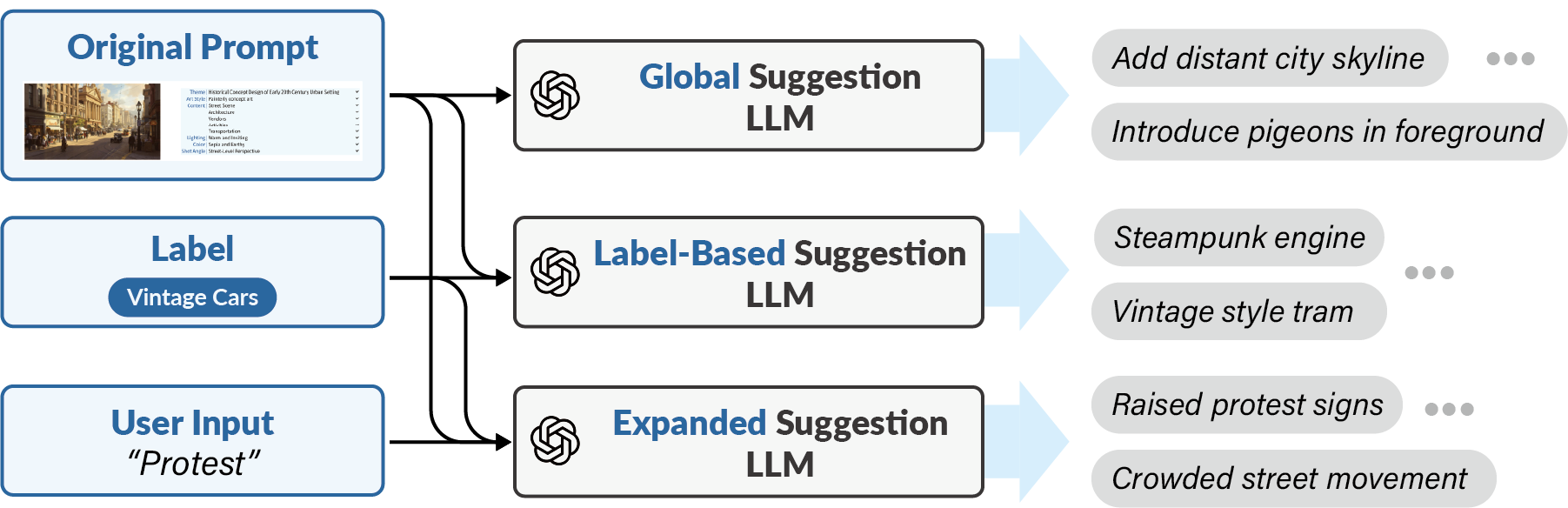}
    \caption{Pipeline for three types of suggestion LLM: Global, Label-based, and Expanded suggestions.}
    \label{fig:suggestion}
\end{figure}

For refinement suggestions (Fig. \ref{fig:suggestion}), GenTune offers three types. Global suggestions use the original prompt to generate five variations focused on content, lighting, and atmosphere. Label-based suggestions generate three refinements and three replacements based on the selected label and its corresponding prompt segment. The expanded suggestions incorporate the user's input, the original prompt, and the selected label (if any) to produce five contextualized options.

%% file: sections/05_summative.tex
\section{SUMMATIVE STUDY}
Our summative study evaluates how GenTune’s two core modules
support professionals in understanding, controlling, and refining generative AI outputs within their workflows.
We conducted two complementary experiments:
\begin{enumerate}
    \item \textbf{A within-subjects experiment simulating real-world generative image refinement workflows.} To benchmark GenTune, we developed a baseline system with a similar UI that reflects common industry workflows. It included: (1) Conversational Image Editing, where designers edited scenes using natural language or reference images via an LLM; and (2) Basic Inpainting, where designers manually selected regions, entered prompts, and applied localized edits using Flux Fill. The baseline excluded GenTune’s two core modules, requiring participants to manually read and iteratively adjust prompts and use inpainting tools for fine-grained control, mirroring typical GenAI workflows. In both conditions, the full structured prompt was displayed (Fig. \ref{fig:ui}-C1). We did not select ChatGPT as a baseline, as it is not well-suited for the complexity of environment design and is not commonly used by professionals in this domain.
    \item \textbf{An open-ended task.} Designers applied GenTune to their own projects involving generative AI tools and compared the experience to their usual workflows.
\end{enumerate}
Our study aimed to answer the following research questions:
\begin{itemize}
    \item \textbf{RQ1:} Can GenTune help designers better interpret the relationship between prompt and image elements?
    \item \textbf{RQ2:} Compared to existing workflows, does GenTune enable more effective and higher-quality refinement?
    \item \textbf{RQ3:} Does GenTune provide greater controllability and better alignment with designers’ expectation during refinement?
\end{itemize}




\subsection{Study Design}
\subsubsection{Procedure}
The study lasted approximately 2 hours and began with a 5-minute briefing.
For the first task, each condition included a 10-minute tutorial, a 10-minute refinement session, and a 5-minute post-task questionnaire. System order and topics were counterbalanced across participants.
The second task began with a 5-minute pre-task interview, followed by a 30-minute design session, where participants explored and applied GenTune in a self-directed manner, they completed a final questionnaire afterwards.
The session concluded with a 15–20-minute post-study interview.


\subsubsection{Task overview}
In the first task, participants completed an image refinement exercise using both the baseline system and GenTune. Each involved one of two assigned design topics with a pre-generated image: (1) The Hanging Gardens of Neo-Babylon or (2) The Floating Monastery of the Himalayas.
Participants completed four rounds of refinement, one global and three local, based on client-style instructions. Local edits specified regions to modify, and participants chose the refinement order freely.
Before each edit, participants identified the prompt corresponding to the element they wished to modify, then selected one of four generated image candidates to continue refining, with up to two iterations per edit. Selection was based on consistency (style, lighting, color, structure, context), aesthetics, and alignment with client intent. Final images were served as references for client communication and future development.
Design topics and refinement instructions were validated by two professional art directors. Figure \ref{fig:example_within} shows an example workflow for the first topic under both conditions.

For the second open-ended task, participants began by describing the workflows of recent environment design projects they had worked on. They then recreated two to three of these projects using GenTune, iteratively refining each image until satisfied, and selecting a final result that aligned with their creative intent and was suitable for client communication.

\subsubsection{Measurements}
The post-condition questionnaire for the first task evaluated three research questions: image–prompt understanding, refinement effectiveness and quality, and overall user experience. It also included a NASA-TLX (Fig. ~\ref{fig:NASA_TLX}) to assess perceived workload. All questions and results are shown in Figure ~\ref{fig:result_within}. Responses were collected using a 7-point Likert scale (1 = strongly disagree, 7 = strongly agree). We adopted a self-report approach, consistent with prior HCI and creativity research~\cite{lubos2024llm,satyanarayan2019critical,palani2022don,son2024genquery}, and analyzed the data using the Wilcoxon signed-rank test~\cite{woolson2005wilcoxon}, 
while NASA-TLX scores were analyzed using paired t-tests.

For the second task, the questionnaire asked participants to rate their preference between GenTune and their previous approach across three core aspects. Responses were recorded on a 7-point Likert scale (1 = strong preference for their original workflow, 7 = strong preference for GenTune). Questions and results are shown in Figure~\ref{fig:comparative}.
We used a one-sample Wilcoxon signed-rank test to assess whether responses significantly differed from the neutral midpoint (4), appropriate for ordinal data from Likert-scale preference questionnaires~\cite{roberson1995analysis, capanu2006testing, taheri2013generalization}. 

In-depth interviews complemented the second open-ended task by offering qualitative insights into how professionals engaged with GenTune in real-world workflows. We focused on how GenTune influenced the understanding of the prompt-image relationship, and how GenTune differed from their previous workflows in terms of interaction flow, editing methods, refinement efficiency and quality. Participants were also asked to elaborate on their questionnaire responses through open-ended questions. All interviews were transcribed and summarized, and user interactions with GenTune were recorded. A lead author with previous experience with environment design conducted the initial coding, collaborating with two additional researchers to identify key themes. The themes were reviewed and finalized through team discussion and consensus.

\subsubsection{Participants}
We recruited 15 professional environment designers with 1–21 years of experience (Mean = 6.60) from industries including games (P1, P4, P6, P11–P13, P17), animation (P3, P5, P14, P21), film (P15–P16), industrial design (P20), and freelancing (P18–P19). They represented over four studios based in Japan, Singapore, and Taiwan, with clients across the EU, Japan, and the US. We also included five design students (P2, P7–P10). Nine participants (P2–P10) had taken part in the earlier formative study and were re-invited via email. The remaining participants were recruited using the same approach. All participants received ~\$30 compensation.


%% file: sections/06_results.tex
\section{RESULTS \& FINDINGS}
In this section, We report findings organized by our three research questions (RQ1–RQ3), combining results from both the controlled experiment and open-ended task.

\begin{figure}
    \centering
    \includegraphics[width=1\linewidth]{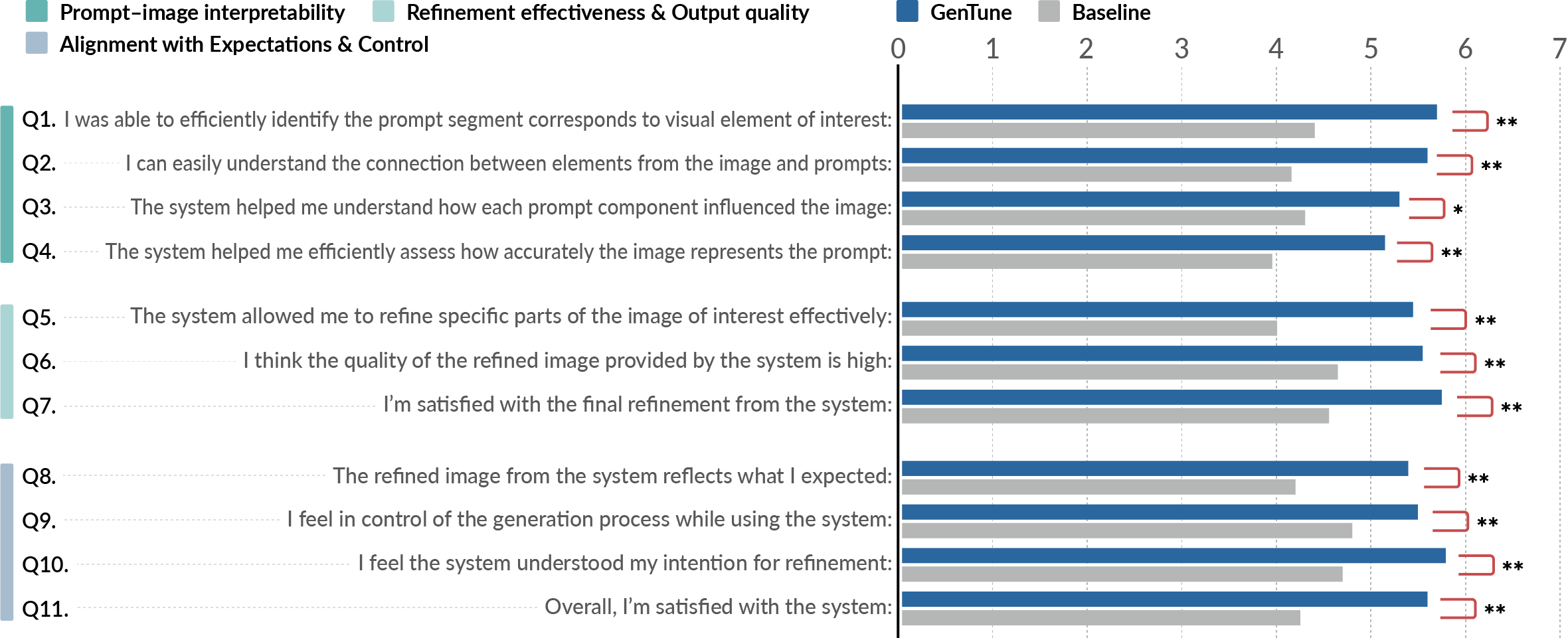}
    \caption{Survey results from the within-subject task. Participants rated Prompt-image interpretability, Refinement effectiveness and quality, and Alignment with expectations and control for both the baseline and GenTune system using a 7-point Likert scale. *: p < .05 and **: p < .01.}
    \label{fig:result_within}
\end{figure}

\begin{figure}[h]
    \centering
    \includegraphics[width=1\linewidth]{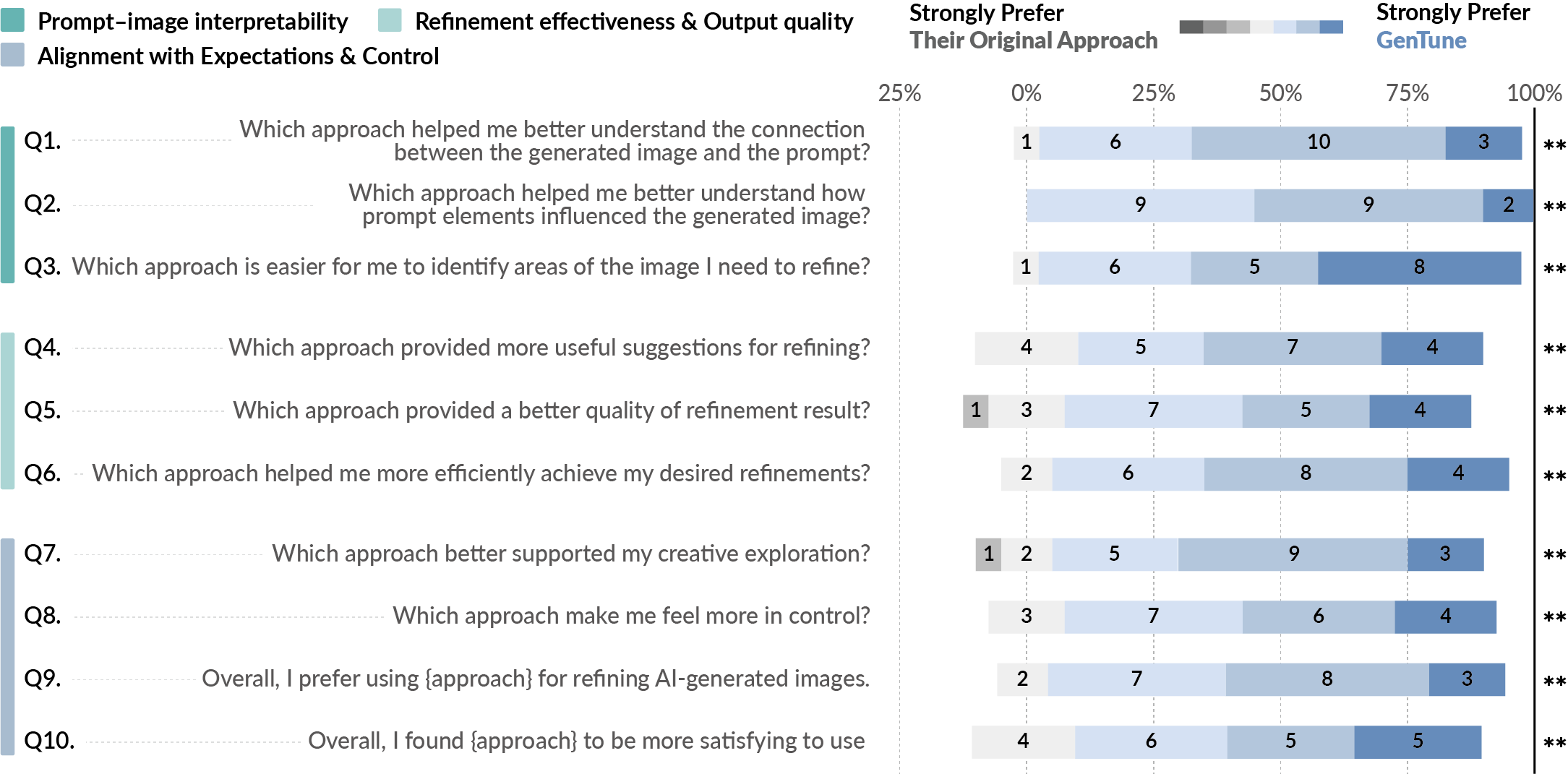}
    \caption{Survey results from the open-ended task. User rated their original approach and GenTune on a 7-point Likert scale across Prompt-image interpretability, Refinement effectiveness and quality, and Alignment with expectations and control. **: p < .01.}
    \label{fig:comparative}
\end{figure}

\subsection{RQ1: Prompt-to-Image Interpretability}
Participants using GenTune demonstrated significantly improved prompt-to-image interpretability.
In the within-subjects study, they found it significantly easier to identify the correspondence between prompts and visual elements, more effectively linked image regions to specific prompts, and better understood how each prompt influenced the image (Fig.\ref{fig:result_within}, Q1–Q3, p < 0.05). In the open-ended task, 95 and 100\% of participants preferred GenTune on these aspects (Fig.\ref{fig:comparative}, Q1–Q3). Many (P2–8, P11–17, P19–21) noted that GenTune saved them from having to read lengthy prompts to find what to change. As P3 explained: “\textit{The highlights clearly show what needs to be changed—I don’t have to dig through the prompt}.” Others found that selecting a label clarified what would be affected (P2–6, P8–13, P16–21): “\textit{The labels are intuitive. Once it’s highlighted, I know what the system will affect. I don’t have to second-guess or worry about unintended changes}” (P6). P5 added, “\textit{The labels help me understand why certain elements appear in the image}.”

However, interpretability decreased in scenes with many similar elements. For example, P11, designing a futuristic plant lab filled with bottles and specimens, found it hard to distinguish between overlapping labels: “\textit{Each candidate seemed plausible, but I couldn’t tell which one was actually correct.}” Another noted “\textit{Sometimes the label is inaccurate, and I have to try several times to get it right}” (P15).

\subsection{RQ2: Refinement Effectiveness and Quality}
\subsubsection{Refinement efficiency and output quality}
Participants rated GenTune significantly more effective for image refinement (Fig.\ref{fig:result_within}, Q5; Mean\_diff = 1.2, p = 0.002), with 90\% expressing a preference for it (Fig.\ref{fig:comparative}, Q6). 
Most reported that semantic-guided refinement accurately targeted the areas they wanted to change (P2–4, P6–10, P12–21) and reduced trial and error. As P12 explained, “\textit{Before, I kept revising prompts because I wasn’t sure what they referred to. With labeled highlights, I know exactly what each part means—no more guesswork.}” 
Some participants found GenTune especially efficient for editing multiple similar elements (P6, P9, P13–14, P17), highlighted its efficiency in adding or replacing elements (P2–3, P6, P10, P12–13, P19). As P9 noted, “\textit{Modifying multiple elements was much faster—GenTune could update all label-related parts at once, previously a slow, manual task in Photoshop.}”

The images refined with GenTune were significantly higher in quality and satisfaction with the final result compared to the baseline (Fig.\ref{fig:result_within}, Q6–Q7, p < 0.01), with 80\% of participants preferring it (Fig.\ref{fig:comparative}, Q5). Many were impressed, as they typically relied on inpainting or manual editing (P3, P5–8, P11–14, P16–17), while GenTune’s semantic-guided prompt refinement delivered better results with minimal structural disruption. As P20 noted, “\textit{GenTune let me control the structure while making precise edits—other tools made unpredictable changes.}”
However, P13 pointed out that even minor structural changes introduced by prompt refinement may be unacceptable to clients: “\textit{Clients often require images to be 100\% consistent}” Similarly, P14 preferred the aesthetic quality of their previous workflow using MidJourney.

80\% of participants preferred GenTune for providing more useful refinement suggestions than their previous workflows (Fig.~\ref{fig:comparative}, Q4). Several noted that these suggestions introduced ideas they hadn’t considered, leading to better outcomes (P3–4, P8–9, P18, P20). As P9 shared, “\textit{When I selected the dragon, the suggestion to make its fire bigger and add a glow to the sword felt very intuitive—I loved it.}” 


\subsubsection{Refinement order}
Since inpainting is not inherently compatible with prompt-based regeneration, just as designers rarely regenerate images after manual edits, participants often planned their strategy in advance, using prompt refinement first, followed by inpainting to avoid overwriting previous changes. As P14 explained, \textit{“I wanted the entire forest of mushrooms to feel more sinister, then fine-tuned the color and shape of individual ones.”} This approach helped maintain consistency and preserve earlier edits.

\subsubsection{Within-subject case comparison}
Table~\ref{tab:within_subject_time} shows the average time and number of refinement iterations for each within-subject task using the baseline and GenTune. Participants took 12.5 minutes and 6.8 iterations with the baseline, compared to 9.23 minutes and 5.3 iterations with GenTune. 

Figure ~\ref{fig:example_within} in Appendix compares the refinement progress of P14 (GenTune) and P16 (Baseline) on topic 1. Both followed the same sequence.
P16 added flowers using global refinement but relied on inpainting for the remaining edits, unsure how to precisely express the changes. He used all iterations, failed two tasks, and was dissatisfied. “\textit{Describing edits through text wasn’t intuitive. Selecting an area felt more natural—more visual thinking.}” In contrast, P14 used GenTune to select areas, assign labels, and apply mixed-mode refinement—completing all four edits in one iteration each using semantic-guided prompts refinement. “\textit{Every step GenTune took precisely matched what I had in mind}” (P14). 

\begin{table}[b]
\centering
\begin{tabular}{|l|c|c|}
\hline
\textbf{Method} & \textbf{Avg. Time Used (min)} & \textbf{Avg. Iterations} \\ 
\hline
Baseline & 12.50 & 6.80 \\
GenTune  & 9.23  & 5.30 \\
\hline
\end{tabular}
\caption{Average time and iterations for the within-subject task using the Baseline and GenTune methods.}
\label{tab:within_subject_time}
\end{table}

\subsection{RQ3: Alignment with Expectations and Control}
Participants found that GenTune’s refinements significantly aligned with their expectations (Fig.\ref{fig:result_within}, Q8; p = 0.002) “\textit{While the output differed from my original idea, it evolved in a different direction—often exceeding my expectations}”(P9).
They rated GenTune as significantly more controllable than the baseline (Fig.\ref{fig:result_within}, Q9; p = 0.003), with 85\% preferring it (Fig.\ref{fig:comparative}, Q8). As P9 noted, “\textit{Extracting the label lets me know exactly what to change. The sense of control comes from the label.}”
This was echoed in open-ended responses: all participants identified label-based selection and editing as GenTune’s most helpful feature, with many (P2, P4, P6–10, P12–13, P16–21) attributing their sense of control to it. As P4 put it, “\textit{I finally felt like I was controlling the AI—other tools feel completely random.}”
However, some participants noted the instablilty in GenTune during refinement “\textit{A building I liked in the previous image disappeared after the modification}” (P21).

Participants were significantly more satisfied with GenTune compared to the baseline (Fig.\ref{fig:result_within}, Q11, p < 0.001) . 90\% preferred GenTune for refining AI-generated images (Fig.\ref{fig:comparative}, Q9) than their previous workflows. All participants expressed interest in using GenTune in future commercial projects. Several noted that it increased their trust in generative design tools (P3–4, P6, P11, P13–16, P21). As P6 stated, “\textit{I can clearly see what the AI will generate, which greatly boosts my confidence in using AI.}” Many participants (P2–5, P7, P9, P11–18, P21) viewed it as a superior tool for communicating with art directors and clients. As P3 shared, “\textit{On a recent project with a tight deadline, the director needed immediate visuals—MidJourney was too unpredictable, but GenTune offered much better control.}” 

85\% of participants preferred GenTune in supporting creative exploration (Fig.\ref{fig:comparative}, Q7) As P17 noted, “\textit{Compared to tools like ComfyUI or Stable Diffusion—where you have to adjust CFG, weights, models, and parameters—GenTune makes it easy. Designers can just focus on the image and the area they want to refine, and it generates exactly what they need.}” This reflects key principles for tools that support creative thinking~\cite{resnick2005design}.


In summary, participants found GenTune to be a more controllable and effective tool for refining AI-generated images. Its intuitive label-based workflow reduced trial and error, while boosting satisfaction, trust, and willingness to adopt it in real-world design workflows.

%% file: sections/07_fieldstudy.tex
\section{FIELD STUDY}
We conducted a three-day field study to evaluate how GenTune supports real-world pre-production workflows. The study focused on the efficiency and quality of the refinement process, and how GenTune helps reduce communication time between designers and stakeholders.

\subsection{Participants, Study Procedure and Evaluation}
We recruited three participants (Table~\ref{table:field}): P4 (Age 27, 4 YoE) from a major game studio known for Metroidvania titles with over 3.6 million copies sold, and P22 and P3 (both Age 27, 3 YoE) from a leading visual effects studio for film and television. These were returning collaborators from our earlier studies, with upcoming projects involving new environment design tasks, and were willing to test GenTune.


We deployed the same GenTune system from the summative study via a web server, adding support for user-uploaded images during generation and refinement. During the study, participants were asked to use GenTune exclusively as their GenAI tool for pre-production tasks, including visual ideation and client communication.

Participants completed a diary study, documenting how they selected elements, issued refinement instructions, and used GenTune in their workflow. A 30-minute post-study interview followed, assessing GenTune’s impact on efficiency, quality, and communication compared to their previous methods. Two researchers independently coded the interviews for thematic analysis.

\begin{table}[h]
\centering
\begin{tabular}{|c|c|c|c|}
\hline
\textbf{ID}  &\textbf{Field} & \textbf{Env. Number} & \textbf{Iterations} \\
\hline
P4 &  Game  & 2 & 2 / 7   \\
\hline
P22 & Visual Effect & 1 & 9 \\
\hline
P3 & Visual Effect & 2 & 7 / 12 \\
\hline
\end{tabular}
\caption{Record of environments completed in the field study. In the \textit{Iterations} column, values in the format x / y represent the number of iterations completed in the first and second environments, respectively.}

\label{table:field}
\end{table}

\begin{figure*}[htbp]
    \centering
    \includegraphics[width=1\linewidth]{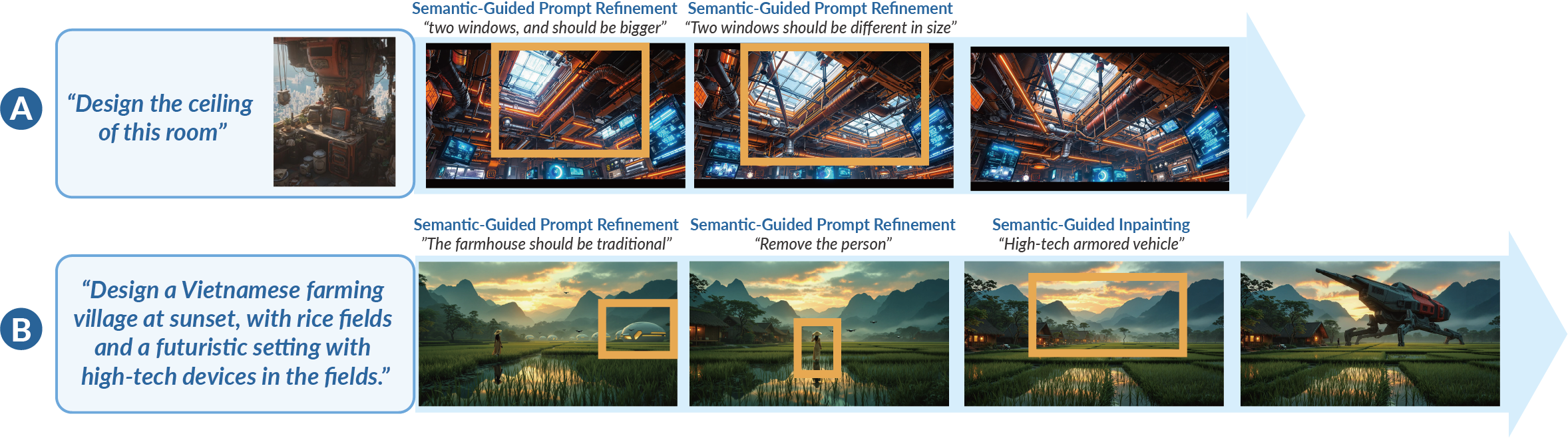}
    \caption{Workflow procedure and results of using GenTune on real-world projects from two designers (P4 and P22). (A) P4 designed a ceiling and room layout based on a reference image using semantic-guided prompt refinement. (B) P22 designed a movie scene blending futuristic elements with Vietnamese rice fields using a combination of semantic-guided prompt refinement and inpainting.}
    \label{fig:field}
\end{figure*}

\subsection{Result}
Table~\ref{table:field} shows the number of environments completed using GenTune, and total iteration counts (each iteration includes four generations or refinements). 

All studios reported notable improvements in efficiency. “\textit{I originally estimated it would take 8 hours to complete, but it only took 2.}” (P22)
Two main factors contributed to this improvement.  First, compared to workflows using ChatGPT and MidJourney, where one refinement often required multiple iterations, GenTune’s traceable prompt structure enabled high-precision refinements. As P3 noted, “\textit{Your tool is more direct and efficient. The local refinement feature significantly reduced the need to go back and forth with Photoshop.}” Second, GenTune accelerated communication between designers and directors. As P22 shared, “\textit{We usually need several meetings to discuss revisions—this time, we completed three rounds of changes in half a day.}”

Figure~\ref{fig:field} presents examples from the field study. P22 was given a two-day deadline to design a movie scene blending futuristic elements with Vietnamese rice fields with the output used directly for 3D scene setup. Starting with the prompt “\textit{Design a Vietnamese farming village at sunset, with rice fields and high-tech devices}”, the client selected one of the generated images and requested two revisions: (1) replace futuraistic buildings with traditional architecture, and (2) remove the person and add a giant armored robot. Using GenTune, the designer completed both changes in 9 iterations, and the final direction was approved. P4 was tasked with designing a ceiling and room layout based on a reference image, with the output serving as a guide for 3D environment development. For the ceiling, she selected the window area and, in just two iterations, achieved a structure suitable for further concept design.

Participants also praised GenTune for enhancing both quality and creativity. As P4 noted, “\textit{I originally thought two skylights were enough, but GenTune suggested making them asymmetrical—and it worked surprisingly well.}” P3 added, “\textit{It often brings out colors or moods I hadn’t considered, which helps give the director an early visual impression.}” Some participants also described unexpected but efficient use cases, “\textit{I can quickly generate props and textures with GenTune and bash them into my draft}” (P3). However, some challenges remained. P22 shared, “\textit{I generated over twenty images to get the giant mech I wanted because the client had already approved the layout. Adding the mech through prompts caused too much structural change.}” Still, P22 acknowledged that GenTune significantly reduced overall revision and communication time.


We are excited to share that following the field study, both studios have continued to use GenTune in their commercial projects, producing over 40 environments to date. This demonstrates GenTune's strong potential for real-world adoption and seamless integration into professional creative workflows.

%% file: sections/08_discussion.tex
\section{DISCUSSION, LIMITATIONS, AND FUTURE WORK}
\subsection{Adapting Generation Approaches to Designers’ Needs}
In the within-subject study, among 80 total refinement actions by 20 participants, 47 involved prompt refining, while 33 used inpainting. They attributed their preference to improvements in quality and consistency. Figure~\ref{fig:seed_better} in Appendix \ref{appendix:A} shows comparisons from our open-ended study, where each image pair was generated with inpainting and with prompt refinement. “\textit{When adding people, the lighting and posture look more consistent and natural with the seed}” (Appendix \ref{appendix:A}, Fig.\ref{fig:seed_better}-A, P17). 
The participants also noted that “\textit{prompt refinement tends to make reasonable changes and offer more possibilities}” (P3). As P16 shared, “\textit{When I added a tram, GenTune also added power lines—I hadn’t even thought of that}”(Appendix \ref{appendix:A}, Fig. ~\ref{fig:seed_better}-B, P16). Most designers were comfortable with the slight variations introduced by prompt refining. This reflects a broader value in environment design: aesthetic coherence and scene plausibility often take precedence over strict pixel-level consistency. This aligns with “The Concept of Coherence in Art”\cite{aschenbrenner2012concept}, which highlights "fittingness" and unity as central to aesthetic experience, even amid minor inconsistencies. 

However, despite offering greater consistency, prompt refinement has notable limitations. For example, P4 attempted to replace a single blackboard with a painting (Appendix \ref{appendix:A}, Fig.\ref{fig:inpainting_better}-A),  but semantic-guided refinement replaced all instances associated with the “blackboard” label.
Similarly, when the prompt modification is broad, fine details may be lost.
In Appendix \ref{appendix:A}, Figure \ref{fig:inpainting_better}-B, P12 used the Izumo Taisha shrine as input reference and wanted to add the traditional shimenawa (straw rope). In the prompt refinement result, while the shrine’s structure was preserved, the torii gate disappeared. In contrast, inpainting resulting in an outcome that better met the designer’s expectations.


We intentionally omitted spatial conditioning controls during refinement, while potentially offering precise structural control, they risk issues such as detail loss, texture degradation, and unintended style shifts. These inconsistencies can disrupt the overall aesthetic coherence, which is critical for designers. 
Figure~\ref{fig:compare_others} in Appendix \ref{appendix:A} compares the results of the GenTune, Flux 1.0 (Depth) and the ChatGPT image editing feature (released 3/25), using the same input: “add prayer flags to the towers”. Both ChatGPT and Flux outputs show significant architectural detail loss, drastic changes in style and texture, and even the disappearance of key elements like the bridge. 


This design implication extends to other creative domains, where different disciplines prioritize different needs and may benefit from distinct generative models. For example, interior designers often require high structural accuracy~\cite{wang2024roomdreaming}, making spatial conditioning controls essential. In contrast, graphic designers may prioritize layout composition~\cite{choi2024creativeconnect}, suggesting that region-based spatial control models may be more appropriate. 

\subsection{Generalizing GenTune-Traceable Control in an Era of Automated Generation}

GenTune’s core concept is a model-agnostic HCI paradigm that enables traceable, element-level control in human–AI collaboration.
Our work addresses a critical challenge emerging from state-of-the-art generative workflows,  where the automated translation of high-level concepts into complex intermediate representations before producing a final output can limit a user's ability to understand, steer, and refine the generative process~\cite{chen2023next}. This is especially true in fast-paced settings like environment design, where designers must rapidly produce variations, making the manual rewriting of prompts impractical.

This challenge applies broadly, from image generation~\cite{wang2025aideation, cai2023designaid} to video, where LLMs act as directors or motion planners~\cite{hong2023direct2v, tan2025mimir, lv2024gpt4motion}, and programming, where LLMs handle planning~\cite{jiang2024self}, multi-step guidance~\cite{han2025multi}, or specification generation~\cite{han2024archcode} prior to code synthesis. 
As these workflows become more automated and multi-step, particularly with the rise of multi-agent systems and MLLMs\cite{zhang2024aflow, dong2024insight, sanwal2025layered, chen2025towards}, the need for understandability and controllability of intermediate outputs becomes increasingly critical. 

Our paradigm introduces traceable links between intermediate representations and final results, enabling users to interpret and revise intermediate steps directly from the output. This traceability significantly enhances designers' sense of control, transparency, trust, and alignment with creative intent in this work. supporting core principles of human-centered AI~\cite{shneiderman2022human, xu2023transitioning, auernhammer2020human}.

GenTune’s implementation can also be directly applied to other complex visual domains that require fine-grained control. For example, P14 successfully used GenTune for character design, effectively controlling individual elements. Similar potential exists in interior design, where elements such as walls, floors, and furniture can be treated independently~\cite{byun2006peculiarity}, and in game UI design, where interface components are inherently structured and easily labeled~\cite{gameui2025}.
Another promising direction is video and animation generation, where evolving visual elements can be labeled with attributes like motion type, speed, and direction. These can be independently adjusted, aligning with established practices in motion design and layered animation~\cite{wang1993layered}, and opening opportunities for traceable control in temporal media.

While our current implementation supports caption-based image inputs, future work could extend the traceable concept to sketches and reference images, enabling users to map generated elements back to specific visual regions, offering even more granular control.

\subsection{Ethical Concern}
As GenAI becomes increasingly embedded in creative workflows~\cite{ko2023}, prior research has raised concerns about its negative impacts, such as the displacement of professional artists~\cite{jiang2023ai, vimpari2023adapt} and the growing pressure for designers to adopt AI tools to stay competitive~\cite{shi2023understanding, vimpari2023adapt}. 

Most of our participants have formal training in digital art and are already expected to integrate GenAI tools into their professional pipelines. GenTune was developed in direct response to the needs and challenges voiced by working designers, and is designed for referencing and facilitating communication, not as a final asset. We see designers as the core creative force, and GenTune aims to augment their workflows by improving control and reducing trial-and-error, allowing them to focus on creative decisions that require their expertise. 

At the same time, we recognize that even positive efficiency gains can contribute to increased client demands, potentially reinforcing the "treadmill" effect. Understanding how AI tools reshape client–designer dynamics is a critical direction for future ethnographic research.


Concerns have also been raised about GenAI reducing group creativity~\cite{doshi2024generative, kumar2025human}. In practice, environment designers “\textit{actively conduct research and curate references, filtering GenAI outputs to integrate their own insights}”, stated by a participant with 13 years of experience (P18). In this sense, systems like GenTune can amplify creative exploration and foster greater diversity through collaboration ~\cite{garcia2024paradox}. However, we acknowledge that overreliance on GenAI may risk homogenization, an important issue for future long-term research on human–AI co-creation.

\subsection{Limitations and Future Work}
\textbf{Study design.} We evaluated GenTune through an open-ended task in which participants applied it to a current or past project, relying primarily on self-reported data. While this approach offers ecological validity, variations in participants' workflows and prior GenAI experience may introduce inconsistencies when comparing GenTune to their original methods. To more rigorously assess refinement effectiveness and output quality, future studies could incorporate a standardized baseline and involve external experts to evaluate and compare refinement results, enabling more insightful analysis. 

\textbf{Refinement order.} While combining inpainting and prompt refinement offers flexibility, it introduces a key limitation: once inpainting is applied, subsequent prompt refinement may regenerate the entire image and overwrite earlier edits. This remains a significant challenge, as designers may struggle to anticipate which method is best suited for a given modification. As P3 noted, “\textit{You need to plan ahead—once you inpaint, it’s difficult to go back and change the overall image.}” GenTune addresses this through a layered system that tracks changes across methods and allows designers to revert to previous versions. Future work could enhance this by introducing a rapid preview system, enabling users to compare refinement outcomes quickly. As models advance, integrating spatial conditioning that merges prompt refinement results with inpainted regions could offer a more seamless solution.

\textbf{Instability of semantic-guided prompt refinement.} Some participants noted the instability of prompt refinement. As P21 remarked, “\textit{Sometimes it accurately modifies only the part I intended, but other times, elements I previously edited disappear.}” This instability stems from the limitations of T2I models, which often struggle to maintain structural and spatial consistency when prompt intent shifts—even with a fixed seed~\cite{bosheah2025challenges, chatterjee2024getting}. Recent work has begun addressing these issues through improved prompt refinement~\cite{manas2024improving}, enhanced spatial understanding~\cite{chatterjee2024getting}, and multi-view consistency~\cite{hollein2024viewdiff}. The development of spatially conditioned control models~\cite{zhang2023adding} also opens new possibilities for more stable and consistent refinements. Future work could integrate these control strategies to strengthen the refinement process further.

\textbf{Label accuracy.} The effectiveness of GenTune’s refinement relies heavily on accurate label selection. If the correct label is missing or unclear, the refinement may not reflect the designer’s intent. Label inaccuracies typically arise from: (1) hallucinations in the T2I model, where elements appear visually but are not captured in prompt-derived labels, and (2) an overabundance of similar or ambiguous labels, making it hard to identify the right one. Beyond improving classification and T2I accuracy, future versions of GenTune could support reverse mapping—highlighting all regions linked to a selected label—to help designers better anticipate which areas will be affected.

%% file: sections/09_conclusion.tex
\section{CONCLUSION}
We present GenTune, a human-centered generative AI system and model-agnostic HCI paradigm that enables traceable, element-level control to improve the understandability and controllability of human–AI collaboration. Designed for environment design workflows, GenTune combines traceable prompts and semantic-guided refinement to help designers better interpret prompt–image relationships and perform more precise, consistent edits. We evaluated GenTune through a summative study with 20 environment designers, including a within-subject experiment and an open-ended design task. Results showed that GenTune significantly improved prompt–image comprehension, refinement quality, efficiency, and user control—receiving strong preference over existing workflows. A follow-up field study in two professional studios further demonstrated GenTune’s potential to enhance refinement efficiency and creative communication in real-world production settings.

%% file: sections/Appendix.tex
\appendix
\section{Appendix}
\label{appendix:A}
\begin{figure*}
    \centering
    \includegraphics[width=0.85\linewidth]{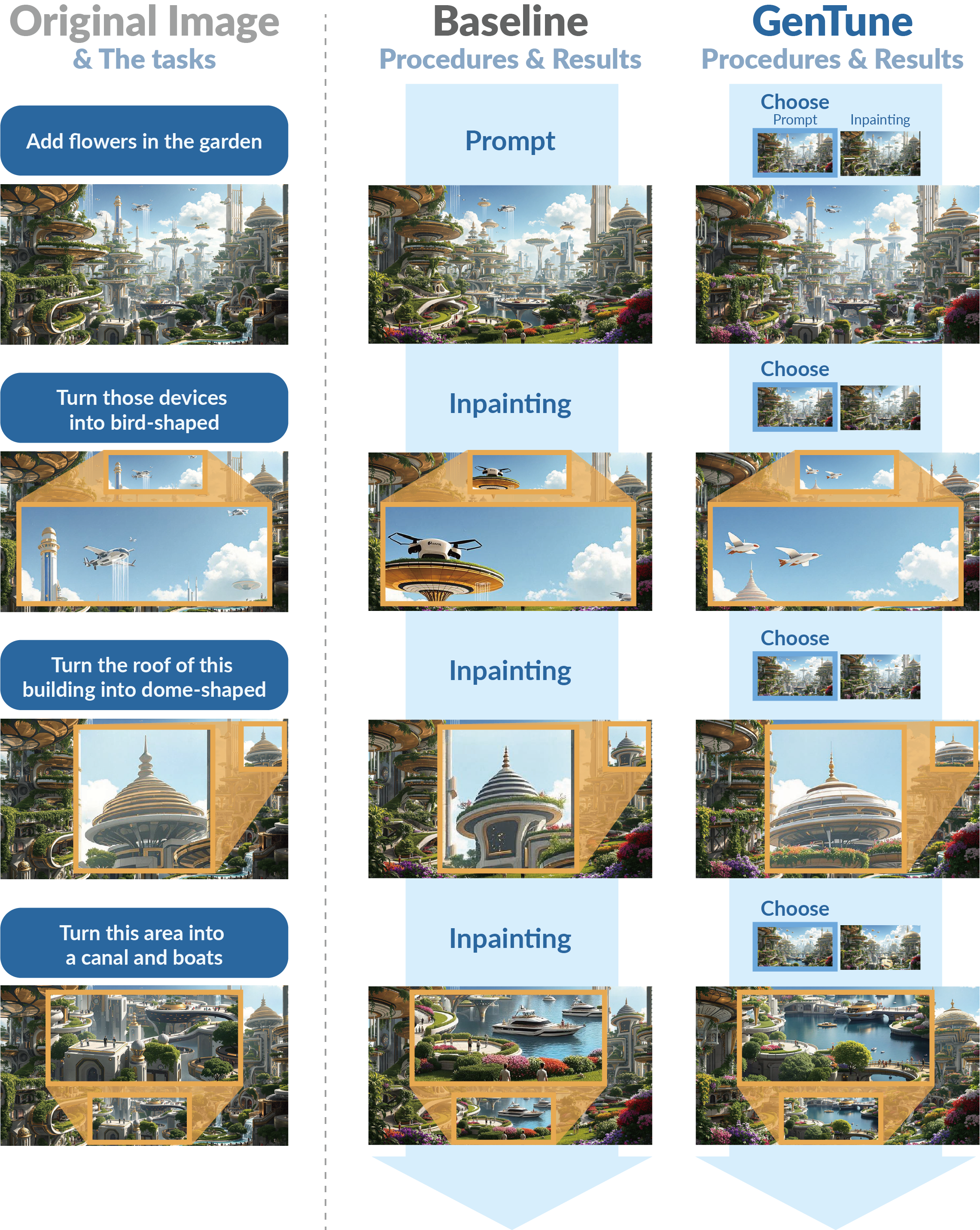}
    \caption{Comparison of refinement workflows between the Baseline and GenTune systems for one of the topics:  The Hanging Gardens of Neo-Babylon in the within-subject task. Participants were shown an initial generated image (left column) and asked to perform four refinement tasks. The center column illustrates the Baseline workflow, which relied on prompt modification and inpainting. The right column shows GenTune’s workflow and the result of user-chosen refinement method between semantic-guided prompt refinement and inpainting.}
    \label{fig:example_within}
\end{figure*}

\begin{figure*}
    \centering
    \includegraphics[width=\linewidth]{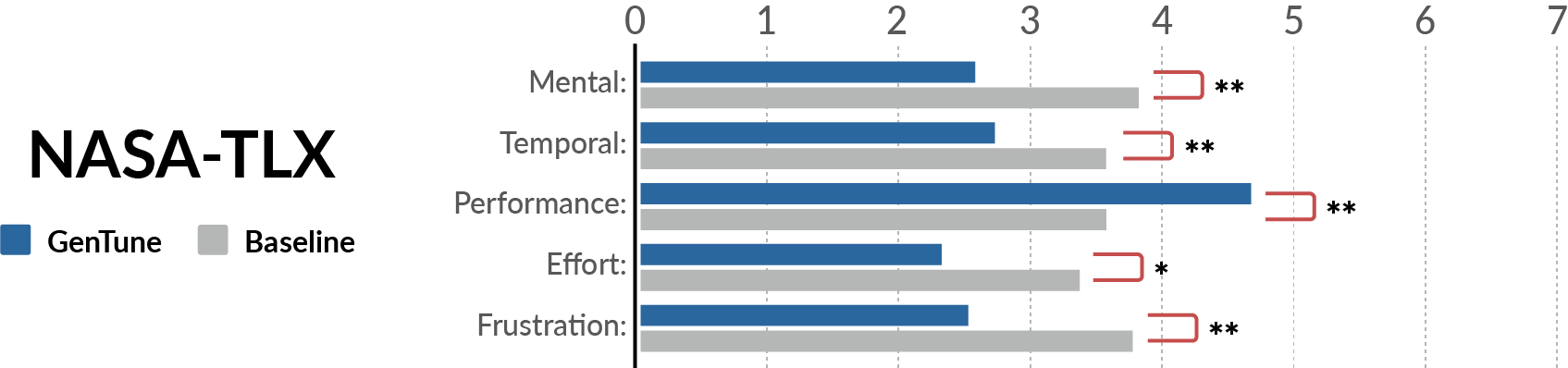}
    \caption{Survey results from the within-subject task. Participants rated the NASA-TLX workload for both the baseline and GenTune system using a 7-point Likert scale. *: p < .05 and **: p < .01.}
    \label{fig:NASA_TLX}
\end{figure*}
\begin{figure*}[htbp]
    \centering
    \includegraphics[width=1\linewidth]{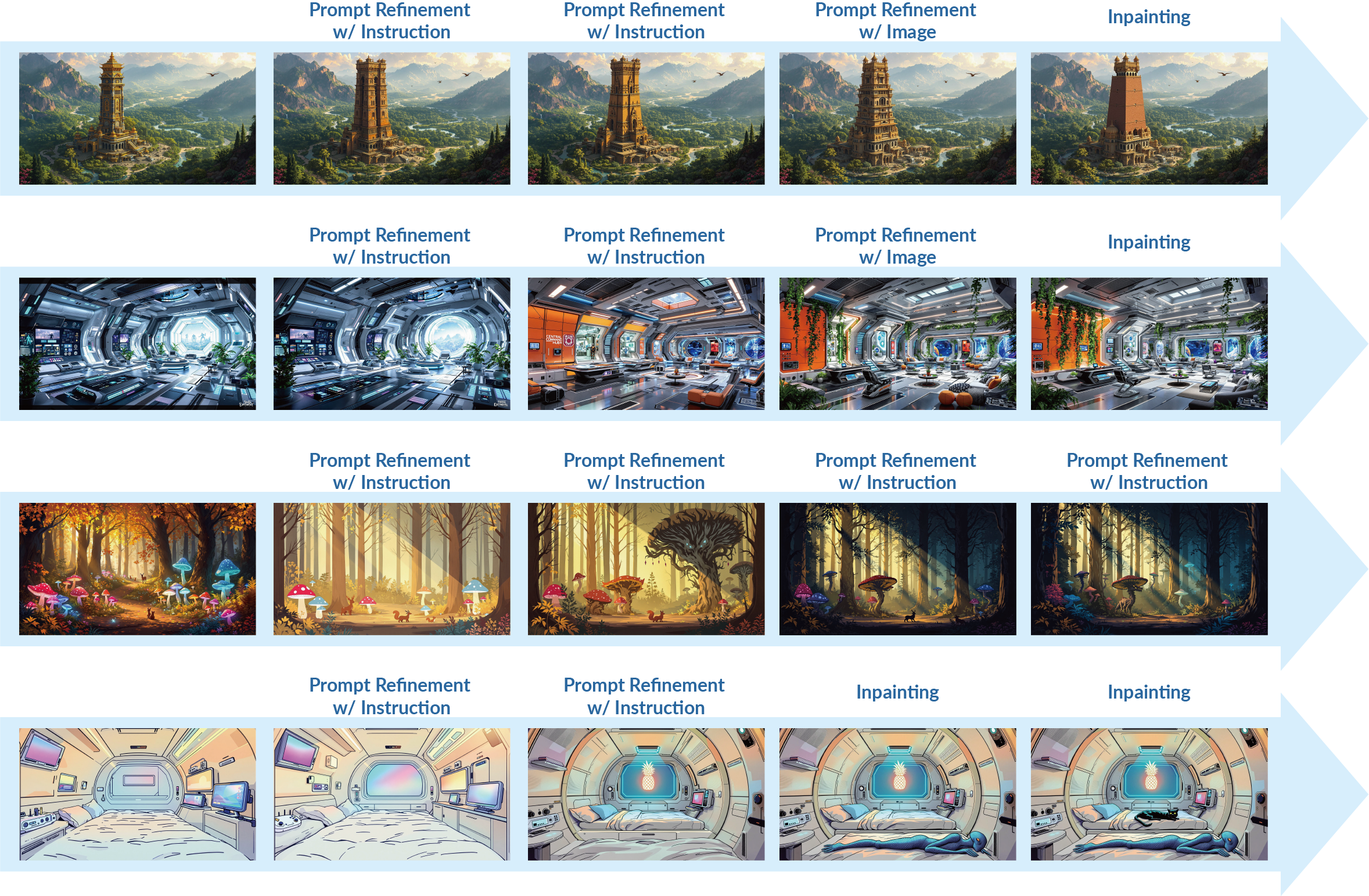}
    \caption{Results from the open-ended task. Each row presents iterative image refinements made by different users using GenTune, including semantic-guided prompt refinement via text instruction or image reference, as well as semantic-guided inpainting.}

    \label{fig:07}
\end{figure*}

\clearpage
\begin{figure*}
    \centering
    \includegraphics[width=1\linewidth]{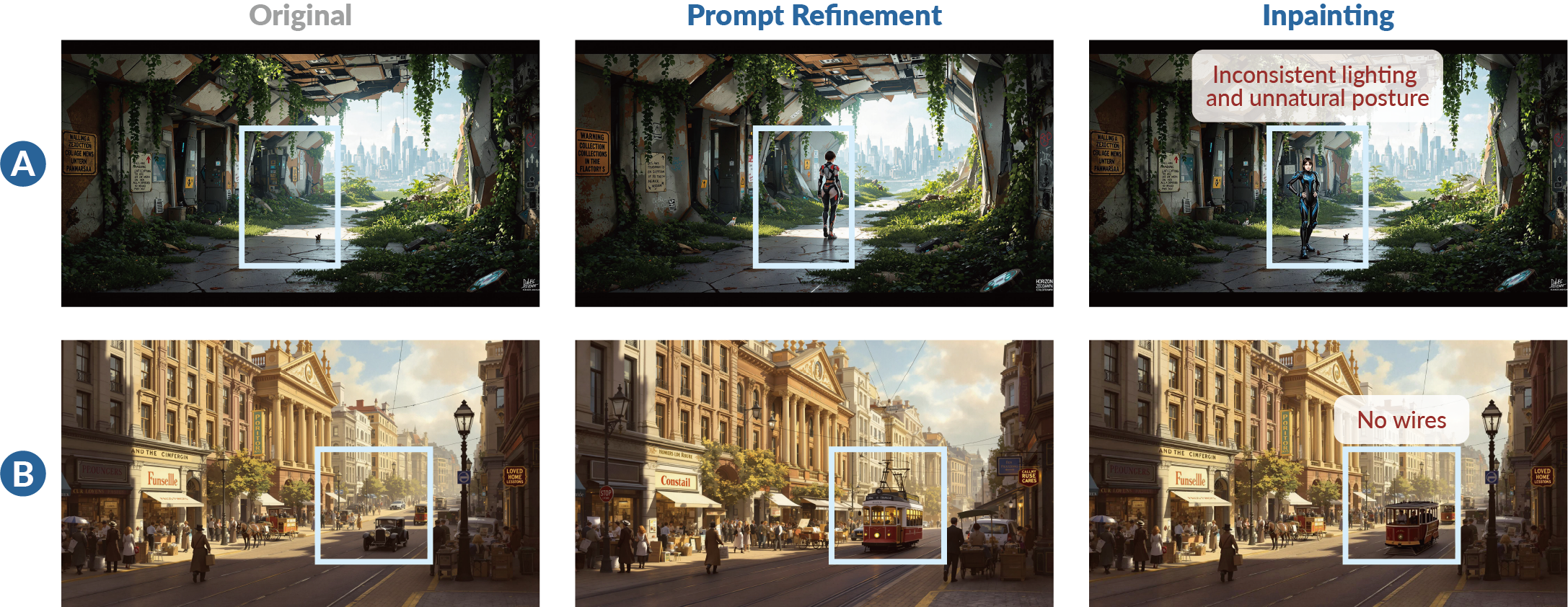}
    \caption{Comparison between semantic-guided prompt refinement with controlled seed (middle) and semantic-guided inpainting (right). In these examples, seed-based refinement better preserves overall image aesthetic coherence.}
    \label{fig:seed_better}
\end{figure*}

\begin{figure*}
    \centering
    \includegraphics[width=1\linewidth]{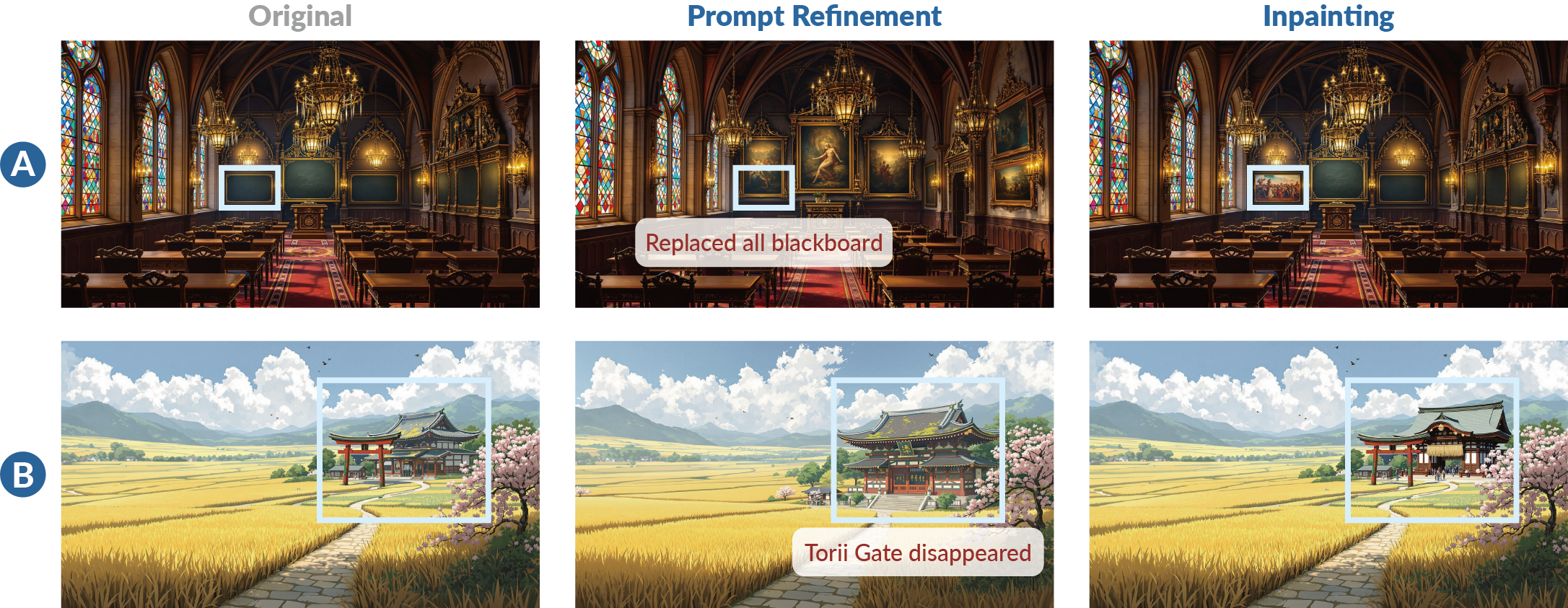}
    \caption{Comparison between semantic-guided prompt refinement with controlled seed (middle) and semantic-guided inpainting (right). In this case, inpainting method provides more precise control.}
    \label{fig:inpainting_better}
\end{figure*}

\begin{figure*}
    \centering
    \includegraphics[width=1\linewidth]{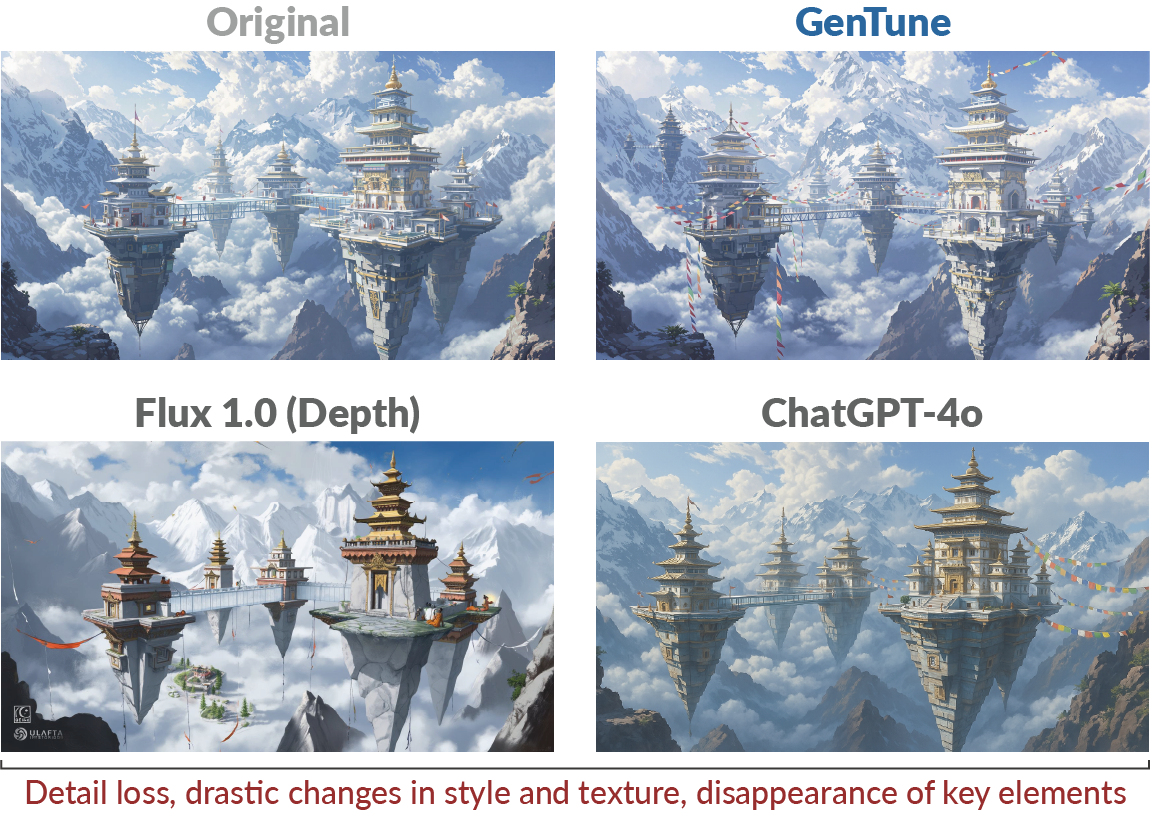}
    \caption{Comparison of image refinement results among GenTune, Flux 1.0 (Depth), and ChatGPT-4o. All methods were given the same instruction: “Add prayer flags on the towers.” GenTune preserves visual coherence, while Flux and ChatGPT exhibit detail loss, drastic changes in style and texture, and even the disappearance of key elements (e.g., the bridge).}
    \label{fig:compare_others}
\end{figure*}